\documentclass[a4paper,11pt]{article}

\usepackage{bbm}
\usepackage[utf8]{inputenc}
\usepackage{placeins}
\usepackage{amssymb,amsmath,amsfonts}
\usepackage[colorlinks=true, linkcolor=blue, citecolor=blue, urlcolor=blue]{hyperref}

\usepackage[dvipsnames]{xcolor}
\usepackage{xcolor}

\usepackage[
    backend=biber,
    style=numeric,
    sorting=none,
    url=true,
    doi=true,
    eprint=true
]{biblatex}
\usepackage{dsfont}
\usepackage{color}
\usepackage{graphicx}
\usepackage{hyphenat}
\usepackage{wrapfig}
\usepackage{empheq}
\usepackage{textcomp}
\usepackage[caption=false]{subfig}
\usepackage{rotating}
\usepackage{wrapfig}
\usepackage{pdfpages}
\usepackage{verbatim}
\usepackage{setspace}
\usepackage{array}
\usepackage{caption}
\usepackage[pdf]{pstricks}
\usepackage{upgreek}
\usepackage{cmll}
\usepackage{latexsym}
\usepackage{braket}
\usepackage{graphicx}
\graphicspath{{grafici}}

\usepackage{epsfig}
\usepackage[left=2.3cm,top=3.2cm,right=2.3cm,bottom=3.2cm,bindingoffset=0cm]{geometry}
\usepackage[titletoc,page]{appendix}
\usepackage{multicol}
\usepackage{youngtab}
\usepackage{setspace}

\usepackage{amsmath}
\usepackage{amsfonts}
\usepackage{amssymb}
\usepackage{mathrsfs}
\usepackage{bm}
\allowdisplaybreaks

\usepackage{tikz}
\usepackage{enumitem}
\usepackage{tensor}
\usepackage{braket}
\usepackage{slashed}
\newcommand{\beq}{\begin{eqnarray}}
\newcommand{\eeq}{\end{eqnarray}}
\newcommand{\bea}{\begin{eqnarray}}
\newcommand{\eea}{\end{eqnarray}}
\newcommand{\be}{\begin{equation}}
\newcommand{\ee}{\end{equation}}

\def\a{\alpha}

\def\nn{\nonumber}

\numberwithin{equation}{section}

\numberwithin{equation}{section}

\begin{document}

\title{
\bf{ Field Theory Models for 
a Holographic Superconductor in 
Two Dimensions} 
\vskip 20pt
} 
\vskip 40pt 
\vskip 40pt   
\author{
 Salvatore Santoro$^{1}$, Roberto Auzzi$^{2}$ and Stefano Bolognesi$^{1}$ 
\\[13pt]
{\em \footnotesize
	$^{(1)}$Department of Physics ``E. Fermi", University of Pisa, and INFN, Sezione di Pisa,}\\[-5pt]
{\em \footnotesize
Largo Pontecorvo, 3, Ed. C, 56127 Pisa, Italy}\\[2pt]
{\em \footnotesize
$^{(2)}$   Dipartimento di Matematica e Fisica,  Universit\`a Cattolica
	del Sacro Cuore,}\\[-5pt]
	{\em \footnotesize
	Via della Garzetta 48, 25133 Brescia, Italy}\\[-5pt]
	{\em \footnotesize
  and  INFN Sezione di Perugia,  Via A. Pascoli, 06123 Perugia, Italy}
\\[2pt]
{ \footnotesize   s.santoro8@studenti.unipi.it, roberto.auzzi@unicatt.it, stefano.bolognesi@unipi.it     }
}

\date{}

\vskip 8pt
\maketitle

\begin{abstract}

We investigate field theory models of holographic superconductors in which the condensation of the order parameter is induced by a Robin boundary condition.
Assuming large-$c$ factorization, we study the phase diagram of 
a  two-dimensional CFT deformed by a relevant double-trace perturbation.
Using modular invariance,  we relate the high- and low-temperature phases,
reproducing analytically the results for the zero-winding sector 
of the holographic model. Moreover, we match the 
near-critical behaviour of the condensate  with an effective Ginzburg--Landau field theory description.
Another important feature of the holographic 
superconductor is the presence 
of vortices that carry fractional magnetic flux. 
We investigate a field theory toy model with similar
properties and interpret it as a fractional Little--Parks effect.

\end{abstract}

\newpage
\tableofcontents


\section{Introduction}

Holographic superconductors provide a useful theoretical laboratory to model strongly coupled systems \cite{Hartnoll:2008vx,Hartnoll:2008kx}.
In two spacetime dimensions, the  Coleman-Mermin-Wagner-Hohenberg (CMWH) theorem \cite{Coleman1973,MerminWagner1966,Hohenberg:1967zz}, forbids spontaneous symmetry breaking.
Superconducting systems in one spatial dimension  have been investigated by many authors in condensed matter physics, see \cite{review-superconductors-1-dim} for a review.
One-dimensional superconductors can be experimentally realized  by considering systems with thickness greater than the coherence length. Superconducting rings do not strictly contradict CMWH theorem,  because superconductivity is always destroyed for sufficiently small wire thickness.

 Holographic superconductors in $1+1$ dimensions are still possible due to the large-$c$ limit \cite{Anninos:2010sq}, which suppresses fluctuations from mean field theory. Holographic superconductors in AdS$_3$ were studied in \cite{Hung:2009qk,Ren:2010ha,Sonner:2014tca,bolognesi}.
In particular,  in the setting studied in \cite{bolognesi} the condensation of
the order parameter was driven by a double trace operator
\cite{witten2001multi,Berkooz:2002ug,multitrace}.

We begin by reviewing the holographic superconductor model recently studied in \cite{bolognesi}. The setup is asymptotically global AdS$_{3}$ and is dual to a $1+1$-dimensional field theory on a circle. The bulk fields are gravity, a $U(1)$ gauge field, and a charged scalar field with mass between the BF bound and zero. The scalar is subject to 
a Robin boundary conditions which is  controlled by the double-trace  deformation 
coupling $\kappa$, which can trigger its condensation and thus drive the dual theory into a superconducting phase.
The corresponding phase diagram in the $T$-$\kappa$ plane has four distinct regions separated by two intersecting transition lines. One line separates the horizonless geometry from the black hole geometry, while the other separates the phase without scalar condensate from the phase where the condensate is non-vanishing. 

These transitions were studied in \cite{bolognesi} from the bulk point of view. Here we address the same structure from the boundary side, with the aim of understanding how much of the holographic phase diagram can be reconstructed directly within field theory. 
In order for a two-dimensional CFT to behave holographically, we need as a condition a large
central charge $c$ \cite{Strominger:1997eq},
large-$c$ factorization \cite{El-Showk:2011yvt,KrausSivaramakrishnanSnively2017} and a sparse light spectrum \cite{Hartman:2014oaa}.
Using these assumptions,
we show that several non-trivial features of the phase diagram can already be captured within a simple effective field theory description. In this framework, one can describe not only the onset of the instability leading to condensation, but also the near-critical behaviour of the condensate itself and its relation to the different thermal regimes of the theory. In particular, in the asymptotic regimes of large and small temperature, the transition line between the condensed and uncondensed phases can be obtained analytically on the field theory side, in agreement with the holographic analysis. More generally, the field theory analysis provides a useful boundary interpretation of the qualitative structure of the phase diagram and of the way the condensate modifies it near the critical region.

Another feature of the holographic model is the presence of vortex solutions with non-zero winding number and magnetic flux. These vortices can exist both in the solitonic phase and in the Black Hole (BH) phase. In the BH phase, they carry an integer magnetic flux hidden behind the horizon.
For vanishing scalar field profile, the BH solution
is given by the BTZ solution \cite{Banados1992,Banados1993}. 
The BH solution realize the Little-Parks periodicity in the dual theory \cite{Montull:2011im,Montull:2012fy}. In the solitonic phase, instead, they display more peculiar features. Most strikingly, their magnetic flux is not quantized and can depend non-trivially on the couplings. Moreover, vortices with different winding number correspond to genuinely distinct solutions, unlike what happens in the black hole phase. One of our aims is to clarify the field theory interpretation of these objects.
To this end, we will present a weakly coupled $1+1$-dimensional toy model that reproduces an analogous fractional-flux effect for the solitonic vortex, as would occur in a hypothetical dual theory. Precisely because this model is weakly coupled, it also helps clarify the relation between this phenomenon and the Little--Parks effect.

The work is organized as follows. In Section~\ref{sec:holo} we review the holographic model. In Section~\ref{sec:field} we discuss the field theory description, the phase transition, and the behaviour of the condensate. In Section~\ref{sec:LP} we introduce the toy model for the fractional Little--Parks effect. We conclude in Section~\ref{sec:conclusion}.
Some technical details are deferred to the appendices.

\section{A holographic superconductor in AdS$_3$}
\label{sec:holo}

In this section we will review the holographic model of a two-dimensional superconductor
\cite{Hung:2009qk,Ren:2010ha,Sonner:2014tca}.
Focusing on the setting where the condensation of the order parameter
is driven by a double-trace perturbation \cite{bolognesi},
we will summarize the
main features of the phase diagram and the
properties of the magnetic vortices. 
A  related study
of the phase diagram 
of hairy rotating black holes
in AdS$_3$ with double-trace
boundary conditions was recently performed
in \cite{Dias:2025uyk,Dias:2026xuy}.
Related soliton solutions 
 have been studied in AdS$_4$ with a multitrace deformation \cite{Dias:2013bwa,Zenoni:2021iiv}.

\subsection{The holographic model}

The holographic model  is  provided by Einstein gravity with a negative cosmological constant, a $U(1)$ gauge field and a charged scalar field with $m^2<0$
\begin{equation}
	S =  \int d^3x \, \sqrt{-g} 
	\left[\frac{1}{16 \pi G_N}\left(
	\mathcal{R} - 2\Lambda\right)
	- \frac{1}{4} F_{\mu\nu}F^{\mu\nu}
- |D_\mu \Phi|^2 - m^2 |\Phi|^2
	\right] ,
	\label{eq:AdS3}
\end{equation}
where
\beq
D_{\mu} \Phi  =\partial_{\mu}  \Phi- i e A_{\mu} \Phi \, ,
\qquad
F_{\mu \nu}  = \partial_{\mu} A_{\nu} - \partial_{\nu} A_{\mu} \, .
\eeq
The AdS radius $L_{\rm AdS}$ is related to the cosmological constant as follows, $L^2_{\rm AdS}=-1/\Lambda$.
The field theory dual is provided by
a 2-dimensional large-$c$  CFT with central charge
\beq
c=\frac{3 L_{\rm AdS}}{2 G_N} \, .
\eeq
From now, we will set the AdS radius $L_{\rm AdS}=1$.
The gravity dual of the vacuum of the CFT on the circle with spatial circumference $L=2 \pi$
is provided by the  global AdS$_{2+1}$ 
spacetime
\begin{equation}
	ds^2 = -(1+r^2)dt^2+\frac{dr^2}{1+r^2}+r^2d\varphi^2 
\end{equation}
A more general solution with cylindrical symmetry
and vanishing angular momentum
can be written in the  form
\begin{align}
	&\Phi=H(r)e^{in\varphi}\ , \qquad A_\mu dx^\mu=\frac{a(r)}{e}d\varphi, \nn \\
	&ds^2 = -(1+r^2)h(r)g(r)dt^2+\frac{h(r)}{g(r)}\frac{dr^2}{1+r^2}+r^2d\varphi^2 \, .
	\label{ansatz}
\end{align}
We set the boundary condition
\beq
\lim_{r\to \infty}h(r)=\lim_{r\to \infty}g(r)=1
\eeq
in order to fix
the asymptotic normalization of the  coordinate $t$.
The integer $n$ corresponds to the winding number of the complex scalar, we refer to the zero winding $n=0$ as the vacuum and $n \neq 0$ for vortex solutions. 
It is useful to introduce the auxiliary function
\beq
q(r)=(1+r^2) \, g(r) \, .
\eeq
The equations of motion are
\beq
\begin{aligned}
& \frac{d}{dr} \left( q \, r  H' \right) = h \,  H \left( \frac{ (n-a)^2}{r} + r \, m^2  \right) , \\
& \frac{d}{dr} \left( \frac{q }{r} \, a' \right) = 2 e^2 \frac{a-n}{r} \, h\,H^2 \, , \\
& \frac{h'}{h} = 8 \pi G \, \left( 2 r {H'}^2 + \frac{ {a'}^2}{e^2  \, r} \right)  \, , \\
& q' = h \left( 2 r - {16 \pi G} \left( \frac{(a-n)^2}{r} + m^2  r \right)  \, H^2 \right) \, . 
\end{aligned}
\label{eom-generiche-2}
\eeq
Eqs. (\ref{eom-generiche-2})
admit both soliton and black hole solutions.

\subsection{Boundary conditions}

The scalar field decays at large $r$ as 
\begin{equation}
	H(r)=\frac{\alpha_H}{r^{\Delta_-}} + \frac{\beta_H}{r^{\Delta_+}} + \dots \qquad \text{with}  \qquad \Delta_\pm = 1 \pm \sqrt{1+m^2}.
\end{equation}
 For Neumann boundary conditions the $\beta_H$ is the source for the operator ${\cal O}_-$ with dimension $\Delta=\Delta_-$ and VEV $\alpha_H$, the opposite for Dirichlet. 
Here we choose  a Robin condition
 for the scalar at the boundary
 \cite{Minces2000}
  so the two coefficients in the decay are related by 
\begin{equation}
	\beta_H = \kappa \ \alpha_H \, .
	\label{double trace condition}
\end{equation}
The mixed condition in
eq. (\ref{double trace condition})
on the CFT side is dual to 
a double trace perturbation
\cite{witten2001multi,Berkooz:2002ug,multitrace}
where  $\kappa$ can be interpreted as a relevant perturbation of the dual theory 
\beq
\delta S =- \kappa \int d^2x \,\, {\cal O}_- {\cal O}_-^{\dagger}
\eeq
of dimension 
\beq
[\kappa]=2\sqrt{1+m^2}=2 (1-\Delta).
\eeq
The double-trace operator can be interpreted as an RG flow between Neumann (UV) and Dirichlet (IR) boundary conditions.
The dimension of the operator ${\cal O}_-$
flows from $\Delta$ in the UV to 
$\Delta_+=2-\Delta$ in the IR \cite{witten2001multi, Hartman2008}.
In order for the deformation to be
relevant and consistent with unitarity,
the dimension of the operator
${\cal O}_-$  must  lie in the window
\beq
0 \leq \Delta < 1 \, .
\eeq
In \cite{bolognesi} 
we studied the case in the
mass range $-1<m^2<-\frac{3}{4}$,
which corresponds 
to $1/2<\Delta<1$.

The behavior of the gauge field at the boundary, in the $A_r=0$ gauge, is
\begin{equation}
	A_\mu = \tilde{\mathfrak{a}}_\mu \log\left(\frac{r}{r_a}\right) + \mathfrak{a}_\mu + \dots \qquad 
\end{equation}  
where $r_a$ is a renormalization scale that can be fixed to $1$. 
Using the standard quantization (the leading term is the source, while the subleading term is the operator) in the dual theory $a_\mu$ is the dynamical gage field while  $\tilde{a}_\mu$ is the external conserved current $\partial^\mu\tilde{a}_\mu=0$ \cite{Jensen2010}.  
In the cylindrical ansatz in eq. (\ref{ansatz}) we take $A_r=A_\varphi=0$
and
\begin{equation}    
   A_t= \frac{a(r)}{e} = 
   \frac{1}{e}
   \left( \tilde{a} \log\left(\frac{r}{r_a}\right) + a_0 + \dots \right)
\end{equation}
In this paper, we will always
restrict to the case of vanishing external current $\tilde{a}$.

\subsection{Black holes}

An interesting class  of solutions of eqs. (\ref{eom-generiche-2})
is provided by black holes.
For a black hole with vanishing 
external current $\tilde{a}=0$
and a generic winding number $n$, the profile function $a(r)$
which solves eq. (\ref{eom-generiche-2})
 is constant, i.e.
\beq
a(r)=n  \, .
\eeq
This property was proved in \cite{bolognesi}, using arguments
similar to the ones used to prove scalar no-hair theorems 
\cite{Bekenstein:1971hc}.

The following conditions 
should be imposed on the horizon
at the horizon $r=r_h$
in order to avoid singularities
\begin{eqnarray}
 \label{H-senza-singo}
 H'(r_h) = H(r_h) \, \frac{h(r_h)}{g'(r_h)} \,   \frac{m^2}{1+r_h^2}
 \, ,
\end{eqnarray}
Black hole solutions have a  temperature  and entropy
\beq T=(1+r_h^2) \, \frac{g'(r_h)}{4 \pi} \, ,
\qquad
S=\frac{ \pi r_h}{2 G}
\eeq
For vanishing scalar hair $H$, a class of  solutions
is provided by the BTZ black hole \cite{Banados1992,Banados1993},
which is recovered by the following profile functions
\beq
h(r)=1, \quad g(r)=\frac{r^2- r_h^2}{r^2+1} 
\eeq
For a given value of the coupling $\kappa$,
a hairy black hole solution exist below 
a critical temperature $T_{c2}$, which can be obtained 
by studying the limit of zero backreaction
on the BTZ solution \cite{bolognesi}.
The result for the critical temperature 
as a function of $\kappa$ is
\begin{equation}
	T_{c2}= \frac{1}{2 \pi} \left[\frac{\Gamma(1-\Delta)}{\Gamma(\Delta -1)}
    \left[
    \frac{\Gamma\left(\frac{\Delta}{2}\right)}{\Gamma\left(1-\frac{\Delta}{2}\right)} \right]^2
    \, \kappa  \right]^{\frac{1}{2(1-\Delta)}}.
	\label{tc ads}
\end{equation}

Black holes realize naturally the periodicity $n\to n+p, \ a(r) \to a(r)+p$ which is valid for the equations and the boundary conditions too. This is interpreted as the periodicity in the Little-Parks effect in the dual theory \cite{Montull:2011im,Montull:2012fy}.

\subsection{Solitons}

Another important class of solutions of eqs. (\ref{eom-generiche-2})
is provided by solitons, which are regular solutions without horizon. At $r=0$ we need to impose  conditions
	\begin{equation}
		H(r) \sim H_0 r^n, \quad a(r)\sim r^2 +a(0), \quad h(0)=g(0).
		\label{boundarysoliton}
	\end{equation}
The solution is regular, apart from the constant $a(0)$ which gives a Dirac string singularity. This possibility was not contemplated in \cite{bolognesi} but here we consider it. 
In order for the Dirac singularity to  be unobservable we must have it quantized \beq
a(0) = k \in \mathbb{Z}.
\eeq
The background, for all physical purposes, is regular at the origin $r=0$. 
  A soliton solution can be put at finite temperature by taking a constant compactification on the Euclidean time.

\begin{figure}
\begin{center}
\includegraphics[scale=0.65]{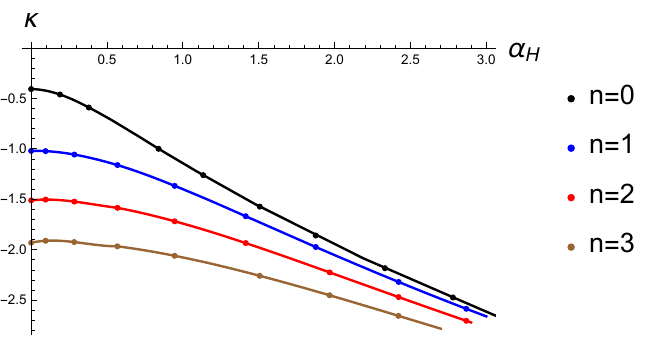} 
\qquad
\includegraphics[scale=0.65]{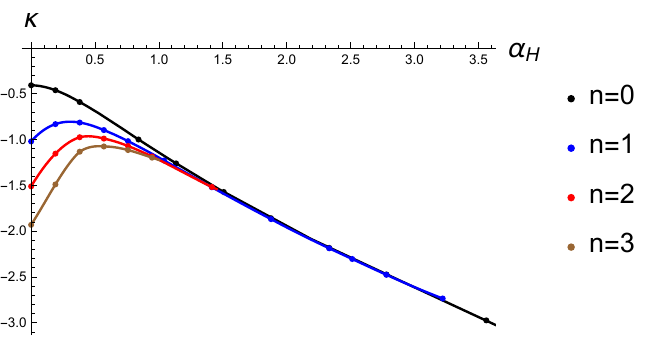} 
\caption{  \small In these plots \cite{bolognesi} we show
$\kappa$ as a function of $\alpha_H$ 
for the vacuum and $n=1,2,3$ vortices. 
Here we set $m^2=-0.9$ and $G=0.1$.
In the left panel we set $e=0$, while in the right panel we consider $e=10$.}
\label{fig-alpha-kappa} 
\end{center}
\end{figure}

For a soliton with $\tilde{a}=0 $ and $a(0)=0$, the relation between the double-trace coupling 
$\kappa$ and the condensate $\a_H$ was investigated numerically in \cite{bolognesi}. 
 In the limit of zero backreaction, it never exists a vortex solution for non-zero winding with the same $\kappa$ of the vacuum. 
We show some results  which include backreaction
in Figure \ref{fig-alpha-kappa}. 
Let us comment on some important features of the numerical solutions
\begin{itemize}
\item
For the vacuum sector $n=0$, the  the condensate $\alpha_H $
is a monotonic function of  the coupling $\kappa$.
This relation does not depend on the gauge coupling $e$.
The superconducting phase transition is second-order.
The critical $\kappa_v$ for the transition can be found
analytically from the solution with zero backreaction.
The result is
\begin{equation}
	\kappa_v=
	\frac{\Gamma(\Delta-1)}{\Gamma(1-\Delta)}
	\left[
	\frac{
	\Gamma\left(1-\frac{\Delta}{2}\right)}
	{\Gamma\left(\frac{\Delta}{2}\right)}
	\right]^2 .
	\label{eq kappa v}
\end{equation}
\item In the vortex sector, instead the condensate
is a monotonic function of $\kappa$ just for $e=0$.
In this case, the superconducting transition is second order,
and the critical coupling $\kappa$ can be found from an
analytical solution in the limit of zero backreaction
\begin{equation}
	\kappa(n)=
	\frac{\Gamma(\Delta-1)}{\Gamma(1-\Delta)}
	\left[
	\frac{
	\Gamma\left(1+\frac{n-\Delta}{2}\right)}
	{\Gamma\left(\frac{n+\Delta}{2}\right)}
	\right]^2 .
	\label{eq kappa soliton}
\end{equation}
For non-vanishing $e$ and $n>0$, there is always another solution
with a higher value of the condensate compared to
the one realized for $\a_H \to 0$.
The solution with higher $\alpha_H$ has always
a lower energy. As a consequence, the superconducting
transition is first order.
\end{itemize}

\begin{figure}
\centering
\includegraphics[width=.6\linewidth]{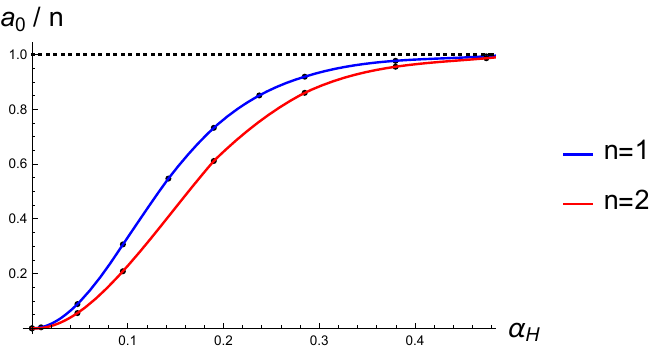}
	\caption{\small  
 Here we plot the normalized flux  $e \, \Phi_B/(2 \pi)=a_0/n$  
 as a function of $\alpha_H$ for the $n=1,2$ vortex solutions 
 \cite{bolognesi}. Here we set $e=10$, $m^2=-0.9$ and $G_N=0.1$.}
	\label{plotflux}
\end{figure}

The total magnetic flux  for the soliton is infinite for non-zero $\tilde{a}$,
 otherwise it is finite and equal to
\begin{equation}
	\Phi_B = \oint A_\mu dx^\mu = \frac{2\pi}{e} (a_0-a(0)).
	\label{flusso}
\end{equation}
  For a vortex in flat space  the flux is quantized 
   for the finite energy condition   $D_\mu \Phi \to 0$.
  The flux is instead not quantized for the AdS vortex solution. For non-zero $\tilde{a}$, we have that the flux
  is infinite. For $\tilde{a}=0$ the flux is 
  proportional  to $a_0$  and it is finite, but not necessarily quantized.
  In our background, we have that the term in the Lagrangian from the covariant derivative is  
  \beq
  H^2 \left(\frac{n-a}{r}\right)^2 \,  \approx \, 
  \a_H^2 \frac{(n-a)^2}{r^{2(2-\sqrt{1+m^2})}}  \, .
  \label{stima-energetica}
  \eeq
  This term does not cause a divergence  
  for $m^2 < 0$.
 We find in fact that $a_0$ is non-quantized and is a dynamical quantity that depends on the parameters in a non-trivial way, see Figure \ref{plotflux}.
The flux $\Phi_B$ can thus be fractional; it depends on the winding $n$ but in a non-trivial dynamical way and can be obtained only by solving the equation of motion for the soliton background. 
 We will provide 
  an explanation of this fractional flux in Section
  \ref{sec:LP}.

\subsection{Phase diagram}

\begin{figure}[h]
	\centering
	\includegraphics[width=.7\linewidth]{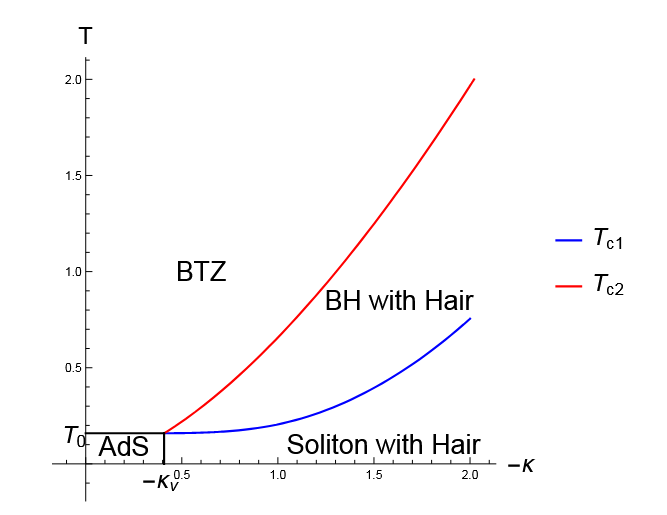}
	\caption{\small  Phase diagram for  $n=0$ with  $m^2=-0.9$
    (corresponding to $\Delta=0.68$)
    and $G_N=0.1$ \cite{bolognesi}. There are four lines departing from a quadruple point dividing the diagram into four distinct phases.}
	\label{diagramma di fase}
\end{figure}

Using holographic renormalization \cite{Skenderis:2002wp,Papadimitriou:2007sj,Caldarelli:2016nni}, in
\cite{bolognesi} the phase diagram
of the system  was investigated.
For the vacuum winding sector $n=0$ the phase diagram 
is independent from the gauge coupling $e$. 
The complete phase diagram  for zero external current 
is shown in Figure \ref{diagramma di fase} 
in the $(-\kappa,T)$ plane. 
for a particular choice of parameters. Now we will describe the various regions of the diagram and the transition lines. 

For sufficiently small $|\kappa|$ there is no scalar condensate.  In this region the two possible solutions are the BTZ black hole and thermal AdS, with a first order 
Hawking-Page transition
\cite{Hawking:1982dh}
between the two geometries
 at $T=T_0=1/2\pi$ . The transition does not depend on $\kappa$, as long as there is no condensate, so it is a horizontal line \cite{Eune2013}. 

Between the hairy and the hairless black hole solutions, there is instead a second order phase transition.  At the phase-transition line $H$ is still vanishing.
As a consequence, we can neglect the backreaction on the gravitational field and study the equation in the BTZ background.
The critical temperature $T_{c2}$ is given in eq. (\ref{tc ads}).
At $T_{c2} \to T_0$ (at $\kappa \to\kappa_v$) this line meets the Hawking-Page phase-transition horizontal line. Later in Section \ref{sec:field}
we will derive the same formula from a purely field  theory computation.
This is one of the original results of the paper.

There is also a first order transition between the soliton with hair and the black hole with hair $T_{c1}$. For this transition,
the backreaction to the metric is important. 
This line was computed numerically. This is, in some sense, a version of the Hawking-Page phase transition in the region of the phase diagram where there is a scalar condensate. This line depends on $\kappa$ as there is dependence on the condensate.  Later in Section \ref{sec:field} we will discuss in more detail the  field  theory interpretation of this phase transition. 

The complete phase diagram  is qualitatively  similar to what is found more generally for other holographic superconductors, even in higher dimensions and with different boundary conditions. An interesting feature 
of our theoretical setting is the presence
of the quadruple point of intersection at $(-\kappa_v,T_0)$ (which in \cite{Nishioka2009} could be obtained only for fine-tuned parameters). 
A similar phase diagram was studied  in \cite{bolognesi}
also for non-vanishing winding $n=1$.

\section{Field Theory model for the phase diagram and the condensate}
\label{sec:field}
In this section we construct an effective CFT description designed to capture the essential ingredients of the holographic model discussed above. In our approach we will not to specify a microscopic CFT, but rather we will isolate some universal properties needed to reproduce the relevant near-critical physics. In particular, we will show that this effective description captures the structure of the phase diagram and the behaviour of the condensate close to the transition lines.

\subsection{The CFT model}
We consider a $1+1$-dimensional CFT containing a scalar primary operator $\mathcal O$ of scaling dimension $\Delta$. The theory is defined in the Euclidean torus $S^1_L\times S^1_\beta$, with coordinates $(\tau,x)\in[0,\beta)\times[0,L)$, where $L$ is the spatial circumference and $\beta$ is the inverse temperature. 
We take $\mathcal O$ to be a single-trace operator
\cite{El-Showk:2011yvt,KrausSivaramakrishnanSnively2017}, namely a light operator whose connected $k$-point functions vanish in the large-$c$ limit for $k>2$. In AdS$_3$/CFT$_2$, large-$c$ is the parameter controlling the semiclassical bulk limit\footnote{In higher dimensional AdS spacetime, the large central charge comes from a large-$N$ limit (the prototypical example for AdS$_5$ is $\mathcal{N}=4$ super Yang Mills theory, where $c\sim N^2$). 
Even when no explicit $N$ is specified,
the factorization property is consistent with the large-$c$ limit \cite{El-Showk:2011yvt}.}.
In this regime, the mean-field description becomes reliable and the Coleman--Mermin--Wagner obstruction \cite{Coleman1973,MerminWagner1966} to spontaneous breaking of a continuous symmetry in $1+1$ dimensions is effectively bypassed at leading order in $1/c$ \cite{Anninos:2010sq}.

For simplicity, throughout this section we treat $\mathcal O$ as a neutral scalar operator. This is sufficient for the purposes of the present effective description of the phase diagram and of the condensate near-criticality. A charged generalization, more directly tailored to the comparison with the AdS superconductor, is certainly of interest, but it is not essential for the considerations developed here and will be left for future work.

We deform the CFT by the
following double-trace perturbation in the Euclidean action
\begin{equation}
	\label{eq:doubletrace}
	\delta S \,=\, \frac{f}{2}\int_0^\beta d\tau \int_0^L dx\; \mathcal O(\tau,x)\,\mathcal O(\tau,x)\,,
\end{equation}
which is interpreted as a double-trace operator in the full theory. 
  We restrict to a relevant double-trace operator $\mathcal{O}^2$, which
corresponds to $\Delta<1$.

We introduce an auxiliary scalar field $\sigma$ through a Hubbard--Stratonovich (HS) transformation \cite{Hartman2008,Gubser2003,Casper2019,Porrati2001,Meng2023} (with scaling dimension $2-\Delta$),
\begin{equation}
	\label{eq:HS}
	\exp\!\left[-\frac{f}{2}\int \mathcal O^2\right]
	=\sqrt{\det\!\left(-\frac{1}{f}\mathbf 1\right)}\int \mathcal D\sigma\;
	\exp\!\left[\frac{1}{2f}\int \sigma^2+\int \sigma\,\mathcal O\right]\,,
\end{equation}
where the overall determinant is fixed by the normalization condition
\begin{equation}
	\sqrt{\det\!\left(-\frac{1}{f}\mathbf 1\right)}\int \mathcal D\sigma\;
	\exp\!\left(\frac{1}{2f}\int \sigma^2\right)=1\,.
\end{equation}

We also introduce an external source $J$ for $\mathcal O$. The generating functional of the undeformed CFT is
\begin{equation}
	Z_{\rm CFT}[J]
	=\int \mathcal D\phi\; e^{-S_{\rm CFT}[\phi]+\int J\,\mathcal O}
	=\left\langle e^{\int J\,\mathcal O}\right\rangle_{0}\,,
\end{equation}
where the subscript $0$ denotes expectation values in the undeformed theory. After the HS transformation, the generating functional of the deformed theory can be written as
\begin{equation}
	\label{Zf}
	Z_f[J]\propto \int \mathcal D\sigma\;
	\exp\!\left(\frac{1}{2f}\int \sigma^2\right)\,
	Z_{\rm CFT}[J+\sigma]\, .
\end{equation}
Equivalently, one may write
\begin{equation}
	Z_f[J]\propto \int \mathcal D\sigma\;
	\exp\!\left(-S_{\rm eff}[\sigma; J]\right)\,,
\end{equation}
with effective action
\begin{equation}
	S_{\rm eff}[\sigma;J]
	= -\frac{1}{2f}\int \sigma^2 - W_{\rm CFT}[J+\sigma]\,
	\qquad \text{where} \qquad
	W_{\rm CFT}[J+\sigma]\equiv \ln Z_{\rm CFT}[J+\sigma]\,.
\end{equation}
In the following we will be interested in the case $J=0$, since our goal is to describe the spontaneous onset of the condensate in the absence of an explicit source.

The functional $W_{\rm CFT}$ generates connected correlators and admits the cumulant expansion
\begin{equation}
	\label{WCFT}
	W_{\rm CFT}[\sigma]
	= W_{\rm CFT}[0]
	+ \sum_{n=1}^{\infty}\frac{1}{n!}\int d^2x_1\cdots d^2x_n\;
	\langle \mathcal O(x_1)\cdots \mathcal O(x_n)\rangle_{0,\,{\rm conn}}\;
	\sigma(x_1)\cdots \sigma(x_n)\,.
\end{equation}
From the equation of motion $\delta S_{\rm eff}/\delta\sigma=0$ one obtains, on-shell,
\begin{equation}
	\label{condizione quasi}
	\frac{1}{f}\,\sigma(\tau,x)+\frac{\delta W_{\rm CFT}[\sigma(\tau,x)]}{\delta\sigma(\tau,x)}=0
	\qquad \Longrightarrow \qquad
	\langle \mathcal O(\tau,x)\rangle = -\frac{1}{f}\,\sigma(\tau,x)\,,
\end{equation}
where the last relation holds at $J=0$ (and it remains valid, with the appropriate shifts, also in the presence of sources).
This makes transparent the source/VEV dictionary on the field theory side: $\sigma$ plays the role of an effective source for $\mathcal O$, while the saddle-point equation relates it directly to $\langle O \rangle$, in close analogy with the AdS relation \eqref{double trace condition}.

Moreover, as shown in \cite{Meng2023}, starting from \eqref{Zf} and performing the rescalings $\tilde\sigma=\sigma+J$ and $\tilde J=J/f$, one finds in the limit $f\to\infty$
\begin{equation}
	e^{W_\infty[\tilde J]}
	=\lim_{f\to\infty}\int \mathcal D\tilde\sigma\;
	\exp\!\left[\frac{1}{2f}\int(\tilde\sigma-f\tilde J)^2 + W_0[\tilde\sigma]\right]
	\propto \int \mathcal D\tilde\sigma\;
	\exp\!\left[-\int \tilde\sigma\,\tilde J + W_0[\tilde\sigma]\right],
\end{equation}
where we dropped the divergent constant factor $e^{\int f\tilde J^{\,2}/2}$. This shows that the generating functionals at $f=0$ (the UV fixed point) and at $f=\infty$ (the IR fixed point) are related by a Legendre transform. In the AdS dual description, this corresponds to the two standard quantizations (Dirichlet and Neumann). In the following, however, we will mainly be interested in the attractive regime \(f<0\), where the double-trace deformation can trigger the condensation of \(\mathcal O\).

\subsection{Large-$c$ effective action and instability criterion}

We now use the assumption that $\mathcal O$ is a single-trace operator of an underlying large-$c$  theory. In the cumulant expansion \eqref{WCFT}, the connected two-point function scales as $c^0$ and dominates over higher connected correlators, which are suppressed at large $c$. Therefore, to leading order in $1/c$ the effective action becomes quadratic and non-local:
\begin{equation}
	\label{Seff non locale}
	S_{\rm eff}[\sigma]
	= -\frac{1}{2}\int d^2x\, d^2y\;
	\sigma(x)\left(\frac{1}{f}\delta^{(2)}(x-y)+G_0(x-y)\right)\sigma(y)+\cdots\,,
\end{equation}
where
\begin{equation}
	G_0(x-y)
	\equiv \langle \mathcal O(x)\mathcal O(y)\rangle_{0,\,{\rm conn}}
	=G_0(\tau-\tau',x-x')\,.
\end{equation}
In momentum space
(see Appendix \ref{app:momentum-space})
this reads
\begin{equation}
	\label{Seff}
	S_{\rm eff}[\sigma]
	= -\frac{1}{2}\sum_{q}\sigma_{-q}\left(\frac{1}{f}+G_0(q)\right)\sigma_q+\cdots\,,
\end{equation}
with $q=(\omega_n,k_m)$, $\omega_n=2\pi n/\beta$, $k_m=2\pi m/L$, and where we introduced the susceptibility
\begin{equation}
	\label{suscettività}
	G_0(q)\equiv \int_{0}^{\beta}\! d\tau \int_{0}^{L}\! dx\;
	e^{-i\omega_n\tau-i k_m x}\,
	\langle \mathcal O(\tau,x)\mathcal O(0,0)\rangle_{0,\,{\rm conn}}\,.
\end{equation}

Stability requires $\frac{1}{f}+G_0(q)<0$ for each mode $q$. Therefore, an instability occurs when the quadratic coefficient changes sign, namely when
\begin{equation}
	\label{condizione critica}
	f_c=-\frac{1}{G_0(q)}<0\,.
\end{equation}
For $|f|<|f_c|$ the effective action has a minimum at $\sigma(x)=0$. When $|f|$ exceeds the critical value, $\sigma(x)=0$ becomes a local maximum and the true minimum is shifted to a configuration with $\sigma(x)\neq 0$, hence $\langle\mathcal O(x)\rangle\neq 0$ via \eqref{condizione quasi}. In the condensed phase one must include a quartic term in $\sigma$ to stabilize the theory at large field values; however, the quadratic analysis is sufficient to locate the phase-transition line.

\subsection{High-temperature limit and critical temperature}

We now compute the critical temperature for the onset of the instability. The first instability typically occurs in the homogeneous mode $q=0$, so we focus on $G_0(0)$.
We start from the high-temperature regime $\beta/L\ll 1$, which is equivalently described by a cylinder with non-compact spatial direction $x\in\mathbb R$ and compact Euclidean time $\tau\sim\tau+\beta$. In this limit one can derive an analytic expression for the transition line.

In a two-dimensional CFT the cylinder can be mapped to the complex plane through the exponential map
\begin{equation}
    z=e^{\frac{2\pi}{\beta}w},\qquad \bar z=e^{\frac{2\pi}{\beta}\bar w},\qquad w=x+i\tau.
\end{equation}
A primary operator with weights $(h,\bar h)$ transforms as
\begin{equation}
	\mathcal O_{\rm cyl}(w,\bar w)
	=\left(\frac{dz}{dw}\right)^{h}\left(\frac{d\bar z}{d\bar w}\right)^{\bar h}\mathcal O_{\rm plane}(z,\bar z), \quad \text{with} \quad
	\frac{dz}{dw}=\frac{2\pi}{\beta}z\,.
\end{equation}
Using the correlator on the plane of the $f=0$ theory
\beq
\langle \mathcal O(z_1,\bar z_1)\mathcal O(z_2,\bar z_2)\rangle_{\mathbb C}
=(z_{12})^{-2h}(\bar z_{12})^{-2\bar h}
\eeq
and
\begin{equation}
    z_{12}=e^{\frac{2\pi}{\beta}w_1}-e^{\frac{2\pi}{\beta}w_2}
=2\,e^{\frac{\pi}{\beta}(w_1+w_2)}\sinh\!\left(\frac{\pi}{\beta}w_{12}\right) \qquad \text{with}
\qquad w_{12}=w_1-w_2,
\end{equation}
one obtains the two-point function on the cylinder
\begin{equation}
	\label{2 punti cft cilindro}
	\langle \mathcal O(w_1,\bar w_1)\mathcal O(w_2,\bar w_2)\rangle_{\beta}
	=\left[\frac{\pi/\beta}{\sinh\!\left(\frac{\pi}{\beta}w_{12}\right)}\right]^{2h}
	\left[\frac{\pi/\beta}{\sinh\!\left(\frac{\pi}{\beta}\bar w_{12}\right)}\right]^{2\bar h}.
\end{equation}
For a scalar primary with $\Delta=h+\bar h$, this becomes (see \cite{BirminghamSachsSolodukhin2003, Solodukhin2005})
\begin{equation}
	\label{eq:cyl2pt}
	\langle \mathcal O(\tau,x)\mathcal O(0,0)\rangle_\beta
	= \left[\frac{\pi/\beta}{\sqrt{\sinh^2\!\left(\frac{\pi x}{\beta}\right)+\sin^2\!\left(\frac{\pi \tau}{\beta}\right)}}\right]^{2\Delta}.
\end{equation}
This is  the full two-point correlator of the $f=0$ theory.
 In the normal phase, where the one-point function vanishes, it coincides with the connected one.
In the condensed phase this is no longer true. However, since the cumulant expansion \eqref{WCFT} is performed around the vanishing-field configuration of the undeformed theory, the quadratic kernel relevant for the following discussion is still determined by the connected two-point function evaluated in that background.

Plugging \eqref{eq:cyl2pt} into \eqref{suscettività} and taking $L\to\infty$, one finds
\begin{equation}
	\label{eq:G0J}
	G_0(0;T)=\left(\frac{\beta}{\pi}\right)^{2-2\Delta}J(\Delta)\,
	\quad \text{with} \quad
	J(\Delta)\equiv \int_{0}^{\pi}\! dv \int_{-\infty}^{\infty}\! du\;
	\frac{1}{\big(\sinh^2 u + \sin^2 v\big)^{\Delta}},
\end{equation}
where $u=\pi x/\beta$ and $v=\pi\tau/\beta$. \\
For $0<\Delta<1$, $J(\Delta)$ is finite and depends only on $\Delta$.

For $A+B\ge 0$ and $0<\Delta<1$ we use the Schwinger representation
\begin{equation}
	(A+B)^{-\Delta}=\frac{1}{\Gamma(\Delta)}\int_0^\infty dt\; t^{\Delta-1}e^{-tA}e^{-tB}.
\end{equation}
Setting $A=\sinh^2 u$ and $B=\sin^2 v$ in \eqref{eq:G0J}, exchanging the order of integration (justified by positivity), and performing the $u$ and $v$ integrals separately, we obtain
\begin{align}
	&\int_{0}^{\pi}\! dv\; e^{-t\sin^2 v}
	= \int_{0}^{\pi} dv\; \exp\!\left[-\frac{t}{2}\big(1-\cos 2v\big)\right]
	= \pi e^{-t/2}\, I_0\!\left(\frac{t}{2}\right), \label{eq:vint}\\
	&\int_{-\infty}^{\infty}\! du\; e^{-t\sinh^2 u}
	= \int_{-\infty}^{\infty} du\; \exp\!\left[-\frac{t}{2}\big(\cosh 2u -1\big)\right]
	= e^{t/2}\, K_0\!\left(\frac{t}{2}\right), \label{eq:uint}
\end{align}
where $I_0$ and $K_0$ are modified Bessel functions. The factors $e^{\pm t/2}$ cancel, and after the change of variables $t=2x$ one finds
\begin{equation}
	\label{eq:Jmellinstart}
	J(\Delta)=\frac{\pi\,2^{\Delta}}{\Gamma(\Delta)}\int_0^\infty dx\; x^{\Delta-1} I_0(x)K_0(x).
\end{equation}
We can now use the Mellin transform formula for $I_0K_0$ (see, for instance, \cite{Bateman1954}), which for $0<\mathrm{Re}(\Delta)<1$ yields
\begin{equation}
	\label{eq:Mellin}
	\int_0^\infty dx\; x^{\Delta-1} I_0(x)K_0(x)
	= 2^{\Delta-2}\,
	\frac{\Gamma\!\left(\frac{\Delta}{2}\right)^{2}\Gamma(1-\Delta)}
	{\Gamma\!\left(1-\frac{\Delta}{2}\right)^{2}},
\end{equation}
and we finally obtain
\begin{equation}
	\label{eq:JGammaVersion}
	G_0(0;T)=\left(\frac{\beta}{\pi}\right)^{2-2\Delta}J(\Delta),
	\qquad
	J(\Delta)=\pi\,2^{2\Delta-2}\,
	\frac{\Gamma\!\left(\frac{\Delta}{2}\right)^{2}\Gamma(1-\Delta)}
	{\Gamma(\Delta)\,\Gamma\!\left(1-\frac{\Delta}{2}\right)^{2}}.
\end{equation}
Combining this with the instability condition \eqref{condizione critica}, we obtain the critical temperature as a function of $f$:
\begin{equation}
	\label{tc toy model}
	T_c(f)=\frac{1}{2\pi}\left[\pi(-f)\,
	\frac{\Gamma\!\left(\frac{\Delta}{2}\right)^{2}\Gamma(1-\Delta)}
	{\Gamma(\Delta)\,\Gamma\!\left(1-\frac{\Delta}{2}\right)^{2}}
	\right]^{\frac{1}{2-2\Delta}}.
\end{equation}
This expression holds in the cylinder limit $L\to\infty$, equivalently $\beta/L\ll 1$. This is also the regime where it is consistent to use the UV scaling dimension for the operator. In intermediate regimes one must instead use the unapproximated expression valid on the torus, where two-point functions are not completely fixed by conformal symmetry (unlike on the plane or on the cylinder) and, in general, one needs to know the spectrum and the OPE data of the theory. The result \eqref{tc toy model} is qualitatively and quantitatively analogous to the holographic expression for the critical temperature \eqref{tc ads}, and to complete the comparison one must identify the precise correspondence between the bulk parameter $\kappa$ and $f$.

\subsection{Modular covariance and the self-dual transition}

In a 1+1-dimensional CFT at finite temperature and finite spatial size, the theory is
defined on the torus \(S^1_L\times S^1_\beta\), with modular parameter
\begin{equation}
\tau_{\rm torus}=i\frac{\beta}{L}.
\end{equation}
For the undeformed CFT the partition function is constrained by modular invariance. In
particular, under the modular \(S\) transformation one has
\begin{equation}
\tau_{\rm torus}\longrightarrow -\frac{1}{\tau_{\rm torus}}
\qquad \text{hence} \qquad
\frac{\beta}{L}\longrightarrow \frac{L}{\beta}.
\end{equation}
Equivalently, the two cycles of the torus are exchanged. For the CFT partition function this
implies
\begin{equation}
Z_{\rm CFT}(\beta,L)=Z_{\rm CFT}(L,\beta),
\end{equation}
up to the usual covariance factors associated with operator insertions. This means that the
low-temperature regime \(\beta\gg L\) and the high-temperature regime \(\beta\ll L\) are
exchanged by modularity \cite{Apolo2022}.

In the deformed theory one must be slightly more careful. The double-trace perturbation
breaks conformal invariance, because its coupling is dimensionful,
\begin{equation}
[f]=2-2\Delta \equiv y_f,
\end{equation}
and therefore introduces the physical scale
\begin{equation}
M_f\sim |f|^{1/y_f}.
\end{equation}
This breaking of conformal invariance does not mean that modular invariance of the torus
path integral is lost. Modular transformations are still large diffeomorphisms of the torus,
and a consistent two-dimensional quantum field theory must be invariant under a relabeling
of the two cycles. Thus, if \(f\) denotes the dimensionful coupling appearing in the action,
one may still write
\begin{equation}
Z_f(\beta,L)=Z_f(L,\beta).
\end{equation}
The non-trivial point is that this is not an invariance at fixed dimensionless coupling.
When the theory is parametrized in terms of dimensionless couplings, modular invariance is
realized as a covariance property on the family of deformed theories.

More explicitly, let us define the dimensionless coupling with respect to the spatial circle,
\begin{equation}
g=fL^{2-2\Delta}.
\end{equation}
After the modular \(S\) transformation the new spatial circle has length \(\beta\). The same
dimensionful coupling \(f\) is therefore described by the transformed dimensionless coupling
\begin{equation}
g'=f\beta^{2-2\Delta}
=
g\left(\frac{\beta}{L}\right)^{2-2\Delta}.
\end{equation}
Thus the modular relation can also be written as
\begin{equation}
Z(\beta,L;g)=Z(L,\beta;g').
\end{equation}
This is the sense in which the double-trace-deformed theory is modular covariant: the
geometric modular invariance of the torus path integral remains true, but in a
parametrization in terms of dimensionless couplings the modular transformation also acts
on the coupling. This is the same logic used for deformed CFTs with dimensionful
couplings, for instance in the \(T\bar T\) case \cite{DattaJiang2018,Aharony:2018bad}.

We now apply this observation to the instability criterion. In the non-condensed phase,
where \(\langle O\rangle=0\), the susceptibility is computed from the undeformed two-point
function on the torus. In the high-temperature regime we already found
\begin{equation}
G_0(0;T)
=
\left(\frac{\beta}{\pi}\right)^{2-2\Delta}J(\Delta),
\qquad
\beta/L\ll 1.
\end{equation}
The instability occurs when
\begin{equation}
\frac{1}{f}+G_0(0)=0.
\end{equation}
Since \(G_0(0;T)\sim \beta^{2-2\Delta}\), the critical temperature satisfies
\begin{equation}
T_c(f)\sim |f|^{\frac{1}{2-2\Delta}}\sim M_f.
\end{equation}
Thus, in the high-temperature branch, condensation starts when the temperature becomes of
the order of the scale generated by the double-trace deformation.

In the opposite regime \(\beta/L\gg 1\), the modular transformation exchanges the
thermal and spatial cycles. Therefore, in the same approximation, the leading dependence
is obtained by replacing the thermal cycle with the spatial one:
\begin{equation}
G_0(0;L)
=
\left(\frac{L}{\pi}\right)^{2-2\Delta}J(\Delta),
\qquad
\beta/L\gg 1.
\end{equation}
This expression is independent of the temperature. The instability condition then gives a
critical value of the coupling,
\begin{equation}
f_d(L)
=
-\frac{1}{G_0(0;L)}
=
-\left(\frac{\pi}{L}\right)^{2-2\Delta}\frac{1}{J(\Delta)}.
	\label{fd toy model}
\end{equation}
Therefore, also in the low-temperature branch, the onset of condensation occurs when the
scale generated by the double-trace deformation becomes comparable to the infrared scale
which cuts off the theory
\begin{equation}
M_{f_d}
\sim |f_d|^{\frac{1}{2-2\Delta}}
\sim \frac{1}{L}.
\end{equation}
In Appendix \ref{m ed n generici} we extend the computation of the critical coupling to
generic momenta \(m\) and \(n\).

The low-temperature endpoint naturally suggests the presence of a quantum critical point.
Indeed \(f_d\) plays the role of a finite-size critical coupling: for \(f>f_d\) the system
remains in the normal phase, while for \(f<f_d\) the double-trace deformation is strong
enough to trigger condensation already in the low-temperature regime. In the holographic
model, the analogous point $\kappa_v$ is the zero-temperature critical value of the double-trace
coupling for which the thermal AdS background becomes unstable toward scalar
condensation. This agreement is one of the non-trivial qualitative matches between the CFT
toy model and the AdS analysis. An analogous quantum critical point is also found in \cite{multitrace}.
One should however keep in mind that, on the torus, the low-temperature
regime is still controlled by the finite scale \(L\), and the corresponding critical coupling
\(f_d\) is therefore not yet the critical point of the infinite-volume theory. The genuinely
scale-invariant quantum critical point is reached only when both cycles are sent to infinity,
namely in the combined limit
\begin{equation}
\beta\to\infty,
\qquad
L\to\infty.
\end{equation}
In that case the torus degenerates into the plane, the finite-size scale disappears, and the
critical point moves to the origin of the \((f,T)\) phase diagram.

The same modular structure also implies the existence of a self-dual transition. The
modular \(S\) transformation has a fixed point when the two cycles have equal size,
\begin{equation}
\beta=L,
\qquad
T_{\rm SD}=\frac{1}{L}.
\end{equation}
At this point the two torus cycles are exchanged into each other. Moreover, the transformed
dimensionless coupling satisfies
\begin{equation}
g'=g,
\end{equation}
so the modular map acts within the same member of the deformed family.

From the CFT point of view, this point separates two different universal regimes: for
\(\beta>L\) the partition function is dominated by the vacuum sector, while for \(\beta<L\)
modularity maps the system to the Cardy regime \cite{Cardy1986}, 
where the asymptotic density of states
controls the thermodynamics. At finite $c$ this is, strictly speaking, only a smooth
crossover, since the partition function is analytic on the compact torus. However, in the
large-\(c\) limit the two competing saddle points become sharply distinct, and the crossover
turns into a genuine thermodynamic transition.

The intersection between this self-dual line and the superconducting instability has a simple
interpretation in terms of scales. Along the high-temperature branch the instability occurs
when \(T_c\sim M_f\), whereas along the low-temperature branch the endpoint is reached
when \(M_f\sim 1/L\). Hence, at the quadruple point $(f_d, T_{SD})$,
\begin{equation}
	T_c \sim \frac{1}{L}\sim M_f .
\end{equation}
In other words, this is the point where the three relevant energy scales of the problem meet:
the temperature, the finite-size gap and the scale generated by the double-trace deformation. Whether the two asymptotic regimes can be reliably extended up to this point will be
discussed in the next subsection.

This self-dual transition is the CFT counterpart of the Hawking--Page transition in
AdS\(_3\)/CFT\(_2\) \cite{Banerjee2018, Kraus2006}. The modular \(S\) transformation exchanges the two cycles of the
boundary torus, while in the bulk this corresponds to exchanging the contractible Euclidean
time circle of the BTZ black hole with the contractible spatial circle of thermal AdS. The
fixed point \(\beta=L\) therefore marks the temperature at which the dominant saddle
changes from thermal AdS to BTZ, so that the self-dual point on the CFT side is precisely
the boundary image of the Hawking--Page transition.

It is important to stress that, in the present discussion, this self-dual transition is a
property of the normal phase. It follows from the modular structure of the undeformed CFT,
together with the modular covariance of the double-trace-deformed family. Once the
condensate contributes non-trivially to the free energy, the same argument no longer fixes
the transition by itself; this is the origin of the possible bending of the self-dual transition
in the condensed phase, as happens on the AdS side.

\subsection{Range of validity of the asymptotic branches}
\label{subsec:range_validity}

The phase structure discussed above was obtained in the two decompactified limits
$\beta/L\ll 1$ and $\beta/L\gg 1$, where the torus reduces to a cylinder and the
two-point function is fixed by conformal symmetry. On a generic torus, instead, the
two-point function is not kinematically fixed and depends on the spectrum and OPE data.
We must therefore clarify under which assumptions the asymptotic expressions used above
can be extended beyond the strict cylinder limits.

A useful result in this direction was obtained in
\cite{KrausSivaramakrishnanSnively2017}. One considers compact, unitary
two-dimensional CFTs at large central charge, introduces a cutoff $\Delta_c$, taken to
infinity after the large-$c$ expansion, and decomposes the spectrum as
\begin{equation}
	\mathcal L=\{A:\Delta_A\le \Delta_c\},\qquad
	\mathcal M=\{A:\Delta_c<\Delta_A\le c/12+\epsilon\},\qquad
	\mathcal H=\{A:\Delta_A>c/12+\epsilon\},
\end{equation}
with $\epsilon>0$ fixed as $c\to\infty$. The assumptions are a sparse light spectrum,
large-$c$ factorization of light correlators, polynomial growth of light correlators in
medium-energy states, and a large-$c$ expansion for the thermal correlator restricted to
light states.

In our normalization, where the modular exchange of the two cycles is not followed by a
rescaling of the torus, the thermal two-point function of a light operator can be written,
up to corrections exponentially small in $\Delta_c$, $\epsilon$ or $c$, as
\begin{equation}
	\label{eq:KSS41_here}
	\langle \mathcal O(x,t)\mathcal O(0,0)\rangle_\beta \approx
	\begin{cases}
	\displaystyle
	\frac{e^{\frac{2\pi\beta}{L}\frac{c}{12}}}{Z(\beta)}
	\sum_{A,B\in\mathcal L}
	e^{-\frac{2\pi\beta}{L}\Delta_A}\,
	e^{\,\frac{2\pi t}{L}(\Delta_A-\Delta_B)}
	e^{\,i\frac{2\pi x}{L}(J_A-J_B)}
	\left|\langle A|\mathcal O|B\rangle_L\right|^2,
	& \beta>L,\\[10pt]
	\displaystyle
	\frac{e^{\frac{2\pi L}{\beta}\frac{c}{12}}}{Z(\beta)}
	\sum_{A,B\in\mathcal L}
	e^{-\frac{2\pi L}{\beta}\Delta_A}\,
	e^{\,\frac{2\pi x}{\beta}(\Delta_A-\Delta_B)}
	e^{\,i\frac{2\pi t}{\beta}(J_A-J_B)}
	\left|\langle A|\mathcal O|B\rangle_\beta\right|^2,
	& \beta<L .
	\end{cases}
\end{equation}
Here $x\sim x+L$ and $t\sim t+\beta$ are the coordinates of the original Euclidean torus,
while $\Delta_A$ and $J_A$ are the scaling dimension and spin of $|A\rangle$. Thus, under
the assumptions of \cite{KrausSivaramakrishnanSnively2017}, the correlator is universal on each side of $\beta=L$ and controlled by light-state data.

In the decompactified limits, the Boltzmann weights project the thermal trace onto the
vacuum in the corresponding channel. For instance, when \(\beta/L\gg 1\), the factor
\(e^{-\frac{2\pi\beta}{L}\Delta_A}\) suppresses all $|A\rangle\neq |0\rangle$ terms, while
\(e^{\frac{2\pi\beta}{L}\frac{c}{12}}/Z(\beta)\to1\) at leading order. The sum over \(B\)
remains and reconstructs the spectral decomposition of
\(\langle 0|\mathcal O(x,t)\mathcal O(0,0)|0\rangle_L\) on the spatial cylinder. In the
strict cylinder limit this is the correlator obtained by the conformal map in
\eqref{eq:cyl2pt}, whose zero mode gives \eqref{eq:G0J}. The high-temperature regime
follows by exchanging the two cycles.

Away from the strict cylinder limits, however, \eqref{eq:KSS41_here} is more general than
the cylinder approximation used above. The sum over \(A\in\mathcal L\) contains
non-vacuum light states, including the state created by \(\mathcal O\) itself and possibly
other light primaries, which need not be negligible close to the self-dual point. Moreover,
even if the identity were the only light primary, the vacuum module would still contain
Virasoro descendants,
\begin{equation}
	L_{-n_1}\cdots L_{-n_k}\,
	\bar L_{-m_1}\cdots \bar L_{-m_\ell}|0\rangle .
\end{equation}
These do not affect the leading Hawking--Page competition: the classical saddle
contribution to \(\log Z\) is of order \(c\), whereas the vacuum character and one-loop
effects are of order \(c^0\). For light correlators, however, the normalized two-point
function is itself of order \(c^0\), so light thermal states and vacuum-module corrections
can change order-one coefficients. For an explicit analysis of such finite-temperature
corrections in large-\(c\) torus two-point functions, see \cite{Pal2022}.
To the best of our knowledge, there is no general CFT argument which removes the
non-vacuum light-sector contributions in \eqref{eq:KSS41_here} and simultaneously extends
the cylinder result in a controlled way up to the self-dual point. The working assumption
made below is instead motivated by the corresponding classical approximation on the AdS
side.

On the AdS side \cite{bolognesi}, one computes the scalar response at tree level on the
dominant classical geometry: thermal AdS for \(\beta>L\) and Euclidean BTZ for
\(\beta<L\). This is the bulk counterpart of keeping only the leading saddle in the thermal
partition function and neglecting quantum fluctuations around it. In particular, the
Virasoro descendants of the vacuum are dual to boundary gravitons associated with the
Brown--Henneaux asymptotic symmetries, and enter as one-loop corrections in the bulk
path integral \cite{Brown:1986nw,Giombi:2008vd,Maloney:2007ud}.\\ In addition, by the standard
state--operator correspondence, a non-vacuum light primary
\(|A\rangle\neq |\mathcal O\rangle\) would be associated with an additional light CFT
operator and, holographically, with an additional light bulk field, absent in the minimal
model considered here. If the state is created by \(\mathcal O\) itself, instead, its thermal
contribution is controlled by a four-point function of \(\mathcal O\); after subtracting the
factorized part, this is a connected finite-\(c\) correction and is not part of the leading
quadratic kernel. Thus, neglecting loops and additional light-sector contributions is the
boundary translation of the classical minimal bulk model, and we apply the same
approximation to the zero-mode susceptibility.

A further indication for the branchwise extension comes directly from the AdS correlators.
The finite-volume BTZ and thermal AdS two-point functions are image sums
\cite{BirminghamSachsSolodukhin2003,Solodukhin2005}, while in the corresponding
cylinder limits they reduce exactly to the CFT expression \eqref{eq:cyl2pt} and to its
low-temperature analogue. In the normalization used above, the Euclidean BTZ contribution
is
\begin{equation}
	\langle \mathcal O(\tau,x)\mathcal O(0,0)\rangle_{\rm BTZ}
	=
	\sum_{n\in\mathbb Z}
	\left[
	\frac{\pi/\beta}{
	\sqrt{
	\sinh^2\!\left(\frac{\pi}{\beta}(x+nL)\right)
	+
	\sin^2\!\left(\frac{\pi \tau}{\beta}\right)
	}
	}
	\right]^{2\Delta}.
\end{equation}
Its zero mode unfolds to the thermal-cylinder susceptibility\footnote{We use the elementary
unfolding of the image sum:
\begin{equation}
	\int_0^L dx \sum_{n\in\mathbb Z} F(t,x+nL)
	=
	\sum_{n\in\mathbb Z}\int_{nL}^{(n+1)L} dx'\,F(t,x')
	=
	\int_{-\infty}^{+\infty}dx'\,F(t,x'), \qquad x'=x+nL \,.
\end{equation}},
\begin{equation}
	\int_0^\beta d\tau\int_0^L dx\,
	\langle \mathcal O(\tau,x)\mathcal O(0,0)\rangle_{\rm BTZ}
	=
	\int_0^\beta d\tau\int_{-\infty}^{+\infty} dx\,
	\left[
	\frac{\pi/\beta}{
	\sqrt{
	\sinh^2\!\left(\frac{\pi x}{\beta}\right)
	+
	\sin^2\!\left(\frac{\pi \tau}{\beta}\right)
	}
	}
	\right]^{2\Delta}.
\end{equation}
Similarly, in the thermal AdS channel the image sum is along Euclidean time, and
\begin{equation}
	\int_0^\beta d\tau\int_0^L dx\,
	\langle \mathcal O(\tau,x)\mathcal O(0,0)\rangle_{\rm AdS}
	=
	\int_{-\infty}^{+\infty} d\tau\int_0^L dx\,
	\left[
	\frac{\pi/L}{
	\sqrt{
	\sin^2\!\left(\frac{\pi x}{L}\right)
	+
	\sinh^2\!\left(\frac{\pi \tau}{L}\right)
	}
	}
	\right]^{2\Delta}.
\end{equation}
Thus, the finite-volume zero-momentum susceptibility computed from the image-summed AdS
correlator on a single classical saddle is exactly equal to the corresponding cylinder
susceptibility. This is precisely the behaviour one would like to reproduce on the CFT side:
the AdS cylinder susceptibility is the same quantity obtained in the asymptotic CFT regimes,
giving \eqref{eq:G0J} in the two branches. The analogous statement is not true for the
full torus correlator. As emphasized in
\cite{BirminghamSachsSolodukhin2003,Solodukhin2005}, the semiclassical AdS prescription
gives image-summed correlators on classical saddles, while a generic CFT torus two-point
function is not obtained by summing images of the cylinder correlator. It depends on the
spectrum and OPE data; in special cases, such as free theories, the answer involves Jacobi
theta functions, which include thermal contributions beyond the vacuum sector. This
mismatch is a well-known open issue in the standard AdS/CFT treatment of finite-volume
thermal correlators on a torus. This suggests that, for the present comparison, the zero-mode susceptibility is a more
natural observable than the pointwise two-point function, since the former is insensitive to
the image sum at the level of a single classical saddle.

The overall logic is therefore the following. We do not claim a formal CFT derivation of
the cylinder approximation up to the self-dual point. Rather, we use the leading
semiclassical approximation suggested by the minimal AdS model: tree-level propagation on
the dominant classical saddle, with no additional light bulk fields and no gravitational
loops. Since in this approximation the finite-volume AdS susceptibility reduces to the same
cylinder susceptibility matched in the asymptotic CFT analysis, we extend the two branches
on this basis. By contrast, the omitted contributions need not be small in genuinely stringy
AdS$_3$ duals, or in CFTs with a dense light sector, additional light fields, or higher-spin
degrees of freedom.

\begin{figure}[t]
	\centering
	\includegraphics[width=0.8\textwidth]{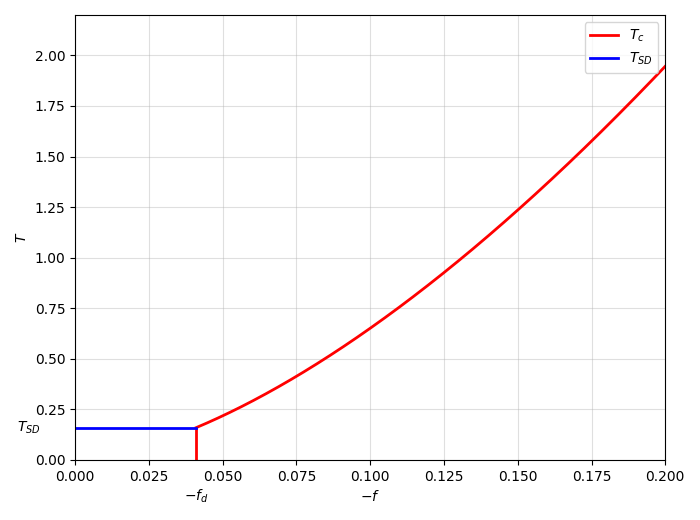}
	\caption{Schematic phase diagram in the \((-f,T)\) plane under the assumptions discussed
	in this subsection, with the scaling dimension fixed to \(\Delta = 1-\sqrt{0.1}\). The superconducting transition separates the normal and condensed
	phases. In the normal phase, the horizontal line \(T_{\rm SD}=1/L\) is the self-dual transition,
	which separates the two modularly related thermal regimes. The continuation of this line
	inside the condensed phase is left undetermined at this stage.}
	\label{fig:phase_diagram_f}
\end{figure}

Under this branchwise assumption, one obtains the phase diagram shown in Figure
\ref{fig:phase_diagram_f}. In the \((-f,T)\) plane, the superconducting transition
separates the normal and condensed phases: at high temperature it is described by
\(T_c(f)\), while at low temperature it terminates at the finite-size critical coupling
\(f_d(L)\). Gluing these regimes gives an effective transition curve \(T_c(f,L)\). In the
normal phase, the self-dual transition at \(T_{\rm SD}=1/L\) separates the two modularly
related thermal regimes. Its intersection with the superconducting line is
\((-f_d,T_{\rm SD})\), which becomes the quadruple point once the continuation of the
self-dual transition into the condensed region, including its bending, is taken into account.

\subsection{Ginzburg-Landau effective action and the condensate}
\label{subsec:landau_condensate}

In order to describe the condensed phase, the quadratic analysis of the previous subsection is not sufficient. Indeed, once the coefficient of the quadratic term changes sign, the configuration $\sigma=0$ becomes unstable, and one must retain the leading nonlinear correction in the effective action in order to stabilize the theory and determine the value of the order parameter. In the large-$c$ framework considered here, this naturally leads to a Ginzburg-Landau (GL) \cite{GinzburgLandau1950} effective action for the auxiliary field $\sigma$. 
Related GL descriptions of holographic superconductors 
have been proposed by various authors, see  for example \cite{Herzog:2008he,Domenech:2010nf,Yin:2015GLTheory,Bu:2021GLEffectiveAction,Natsuume:2025DualGL,Natsuume:2024DualGLFiniteCoupling}. 
The perspective adopted here is different. Rather than deriving the GL functional from
the bulk equations or postulating it as a phenomenological description of the holographic
phase transition, we obtain it directly from the boundary large-$c$ effective action generated
by the double-trace deformation. In this formulation the quadratic coefficient is determined
by the CFT two-point function, while the quartic coupling is controlled by the connected
four-point function of the operator $\mathcal O$. Thus the GL description arises as the
near-critical expansion of the boundary effective theory.

Expanding the connected generating functional $W_{\rm CFT}[\sigma]$ \eqref{WCFT} beyond quadratic order, and restricting for simplicity to the homogeneous mode relevant for the onset of the transition, one obtains the local Euclidean effective action
\begin{equation}
	S_{\rm eff}[\sigma]
	=
	- W_{\rm CFT}[0]
	-\int d\tau\,dx\,
	\left[
	\frac{1}{2}c_\tau (\partial_\tau \sigma)^2
	+\frac{1}{2}c_x (\partial_x \sigma)^2
	+\frac{1}{2}r(T,L)\,\sigma^2
	-\frac{1}{4!}u(T,L)\,\sigma^4
	\right] .
	\label{eq:Seff_complete}
\end{equation}
Here the quadratic coefficient is
\begin{equation}
	r(T,L)=\frac{1}{f}+G_0(0;T,L),
	\label{eq:r_def}
\end{equation}
while the quartic coupling is determined by the connected four-point function of the undeformed CFT,
\begin{equation}
	u(T,L)=-\int d^2y\,d^2w\,d^2z\;
	\langle \mathcal O(0)\mathcal O(y)\mathcal O(w)\mathcal O(z)\rangle_{0,\mathrm{conn}} .
	\label{eq:u_def}
\end{equation}
The coefficients \(c_\tau\) and \(c_x\) encode the leading momentum dependence of the susceptibility, defined through the small-momentum expansion
\begin{equation}
    G_0(q)=G_0(0)+c_\tau \omega_n^2 + c_x k_m^2 + O(q^4).
    \label{G espanso}
\end{equation}
Their explicit expression will be computed in Appendix \ref{app:coefficients}. \\
When this expansion is inserted into the quadratic effective action \eqref{Seff} in momentum space, the terms proportional to \(\omega_n^2\) and \(k_m^2\) generate the local derivative terms in position space. Indeed, upon Fourier transforming back, one has
\begin{equation}
    \sum_q \sigma_{-q}\,\omega_n^2\,\sigma_q
    =
    \int_0^\beta d\tau \int_0^L dx\, (\partial_\tau \sigma)^2,
    \qquad
    \sum_q \sigma_{-q}\,k_m^2\,\sigma_q
    =
    \int_0^\beta d\tau \int_0^L dx\, (\partial_x \sigma)^2.
\end{equation}
For the homogeneous and static configurations considered in the following, however, these derivative terms do not contribute, and we may therefore set \(\partial_\tau \sigma=\partial_x \sigma=0\).\\
Moreover, in the present neutral truncation, a cubic term would in principle be allowed. We nevertheless restrict the Ginzburg-Landau effective action to be even in \(\sigma\), keeping in mind the charged generalization relevant for comparison with the AdS superconductor, where cubic terms are forbidden by the \(U(1)\) transformation properties of the order parameter.

For a uniform configuration, the effective action reduces to an effective potential density,
\begin{equation}
	V_{\rm eff}(\sigma)
	=
	-\frac{1}{2}r(T,L)\,\sigma^2+\frac{1}{4!}u(T,L)\,\sigma^4 .
	\label{eq:Veff_sigma}
\end{equation}
Its extrema satisfy
\begin{equation}
	\frac{\partial V_{\rm eff}}{\partial \sigma}
	=
	-r(T,L)\,\sigma+\frac{u(T,L)}{6}\sigma^3
	=0 ,
\end{equation}
and therefore the solutions are
\begin{equation}
	\sigma=0,
	\qquad
	\sigma^2=\frac{6\,r(T,L)}{u(T,L)} .
\end{equation}
Equivalently, with the sign convention used in \eqref{eq:Seff_complete}, one may write the stationary solutions as
\begin{equation}
	\sigma=
	\begin{cases}
	0, & r(T,L)<0,\\[4pt]
	\pm\sqrt{\dfrac{6\,r(T,L)}{u(T,L)}}, & r(T,L)>0.
	\end{cases}
	\label{eq:sigma_minima}
\end{equation}
Thus the normal phase and the condensed phase are distinguished by the sign of $r(T,L)$, namely by whether the instability condition has been crossed.

Using the relation derived earlier between the Hubbard--Stratonovich field and the operator expectation value,
\begin{equation}
	\langle \mathcal O\rangle = -\frac{\sigma}{f},
	\label{eq:O_sigma_relation}
\end{equation}
we immediately obtain the condensate. In the non-condensed phase one has
\begin{equation}
	\langle \mathcal O\rangle =0 ,
	\qquad r(T,L)<0 ,
\end{equation}
whereas in the condensed phase
\begin{equation}
	\langle \mathcal O\rangle
	=
	 \frac{1}{|f|}
	\sqrt{\frac{6\,r(T,L)}{u(T,L)}} ,
	\qquad r(T,L)>0 .
	\label{eq:condensate_general}
\end{equation}
This provides the general mean-field expression for the order parameter on the torus.

It is convenient to rewrite this result separately in the two asymptotic regimes discussed above.

In the high-temperature regime $\beta/L\ll 1$, using
\begin{equation}
	G_0(0;T)=A(\Delta)\,\beta^{2-2\Delta},
	\qquad
	u(T)=B_\beta(\Delta)\,\beta^{6-4\Delta},
\end{equation}
the homogeneous minimum of the effective action gives
\begin{equation}
	\langle \mathcal O\rangle_{\rm high\,T}
	=
	\sqrt{\frac{6\,G_0(0;T)}{u(T)\,|f|^2}
	\left[
	1-\left(\frac{T}{T_c}\right)^{2-2\Delta}
	\right]}
	=
	\frac{T^{2-\Delta}}{|f|}
	\sqrt{\frac{6\,A(\Delta)}{B_\beta(\Delta)}
	\left[
	1-\left(\frac{T}{T_c}\right)^{2-2\Delta}
	\right]},
	\qquad T<T_c .
	\label{eq:cond_highTT}
\end{equation}
Close to the critical temperature, one can expand
\begin{equation}
	1-\left(\frac{T}{T_c}\right)^{2-2\Delta}
	\simeq
	(2-2\Delta)\left(1-\frac{T}{T_c}\right).
\end{equation}
Moreover, within the same near-critical approximation one may replace
\(T^{2-\Delta}\) by \(T_c^{\,2-\Delta}\) in the prefactor. It follows that the condensate
vanishes with the standard mean-field critical exponent,
\begin{equation}
	\langle \mathcal O\rangle_{\rm high\,T}
	\propto
	\left(1-\frac{T}{T_c}\right)^{1/2},
	\qquad T\to T_c^- ,
\end{equation}
as expected from a Ginzburg-Landau effective description of a second-order
phase transition.

In the opposite regime $\beta/L\gg 1$, modular covariance fixes the corresponding
scaling with the finite-size scale \(L\). At quadratic order this gives the same universal
coefficient \(A(\Delta)\), because the instability is controlled by the two-point function.
At quartic order, however, the coefficient is determined by the integrated connected
four-point function of \(\mathcal O\), and is therefore not fixed by conformal symmetry
alone. In an effective description of the two asymptotic branches we will therefore allow
for an independent non-universal coefficient in the low-temperature regime:
\begin{equation}
	G_0(0;L)=A(\Delta)\,L^{2-2\Delta},
	\qquad
	u(L)=B_L(\Delta)\,L^{6-4\Delta}.
\end{equation}
This does not mean that modularity of the full theory is lost. Indeed modular covariance is a statement about the full torus partition function, and need not be
manifest term by term in a truncated Ginzburg-Landau expansion around a given saddle. Moreover,
the semiclassical approximation used here keeps only the two dominant saddles in the
asymptotic high- and low-temperature regimes. Subleading saddles and loop
corrections, although irrelevant for the leading Hawking--Page competition, may contribute
to the full modular completion and to non-universal coefficients such as the quartic GL
coupling.

The transition is controlled by the critical coupling $f_d$ \eqref{fd toy model}. The
condensate in this low-temperature branch is
\begin{equation}
	\langle \mathcal O\rangle_{\rm low\,T}
	=
	\frac{L^{2\Delta-3}}{|f|^{3/2}}
	\sqrt{\frac{6}{B_L(\Delta)}
	\left(\frac{|f|}{|f_d|}-1\right)},
	\qquad |f|>|f_d| .
	\label{eq:cond_lowT}
\end{equation}
Close to the transition, one may approximate $|f|^{-3/2} \simeq |f_d|^{-3/2}$, so that
\begin{equation}
	\langle \mathcal O\rangle_{\rm low\,T}
	\simeq a_L\left(\frac{|f|}{|f_d|}-1\right)^{1/2} \qquad \text{with} \qquad a_L=\frac{L^{2\Delta-3}}{|f_d|^{3/2}}
	\sqrt{\frac{6}{B_L(\Delta)}} 	=
	L^{-\Delta}
	\sqrt{\frac{6 \, A(\Delta)^3}{B_L(\Delta)}} \,.
    \label{behaviour bassa T}
\end{equation}

In the high-temperature branch the same expansion can be performed by replacing the
finite-size scale \(L^{-1}\) with the temperature \(T\). Equivalently, one introduces the
temperature-dependent critical coupling
\begin{equation}
	f_c(T)=-\frac{1}{G_0(0;T)}
	=
	-\frac{T^{2-2\Delta}}{A(\Delta)}.
\end{equation}
The condensate can then be written as
\begin{equation}
	\langle \mathcal O\rangle_{\rm high\,T}
	=
	\frac{T^{3-2\Delta}}{|f|^{3/2}}
	\sqrt{\frac{6}{B_\beta(\Delta)}
	\left(\frac{|f|}{|f_c|}-1\right)},
	\qquad |f|>|f_c|.
	\label{eq:cond_highT}
\end{equation}
Close to the transition, one may approximate $|f|^{-3/2} \simeq |f_c|^{-3/2}$, so that
\begin{equation}
	\langle \mathcal O\rangle_{\rm high\,T}
	\simeq a_\beta\left(\frac{|f|}{|f_c|}-1\right)^{1/2} \qquad \text{with} \qquad a_\beta=\frac{T^{3-2\Delta}}{|f_c|^{3/2}}
	\sqrt{\frac{6}{B_\beta(\Delta)}} 	=
	T^{\Delta}
	\sqrt{\frac{6 \, A(\Delta)^3}{B_\beta(\Delta)}} \,.
    \label{behaviour alta T}
\end{equation}


Let us stress that the only non-universal inputs in the condensates are the coefficients
$B_\beta(\Delta)$ and $B_L(\Delta)$. These are not fixed by conformal symmetry or by the
two-point function, since they depend on higher connected correlators and therefore on the
specific CFT under consideration. For this reason we will later extract them from
the comparison with the holographic condensate.

\subsection{Bending of the self-dual transition}
\label{subsec:selfdual_bending}

In the normal phase, modularity fixes the self-dual transition at
\(T_{\rm SD}=1/L\). In the condensed phase, the condensate contributes to the free energy
and can shift the point at which the high-\(T\) and low-\(T\) branches exchange dominance.
As discussed above, this should not be interpreted as a breaking of modularity of the full
theory, but rather as a possible effect of the branchwise Ginzburg--Landau truncation, in
which the quartic coefficients in the two asymptotic regimes need not coincide.

For homogeneous and static configurations, the condensate contribution to the free energy is
\begin{equation}
    F_{\rm cond}
    =
    \frac{1}{\beta}\left(S_{\rm eff}[\sigma]+W_{\rm CFT}[0]\right)
    =
    -\frac{3}{2} L \frac{r^2(T,L)}{u(T,L)} .
    \label{energia libera cond}
\end{equation}
Using the near-critical GL expansion in the high-temperature branch, the free energy can
be written as
\begin{equation}
F_T(f,L,T)=
\begin{cases}
\displaystyle
-\frac{3A(\Delta)^2}{2B_\beta(\Delta)}
L\,T^2
\left(\frac{f}{f_c(T)}-1\right)^2
-\dfrac{\pi c}{6}LT^2,
& f>f_c(T),\\[10pt]
\displaystyle
-\dfrac{\pi c}{6}LT^2,
& f<f_c(T).
\end{cases}
\label{eq:FT_bending}
\end{equation}
Similarly, in the low-temperature branch one finds
\begin{equation}
F_L(f,L,T)=
\begin{cases}
\displaystyle
-\frac{3A(\Delta)^2}{2B_L(\Delta)}
\frac{1}{L}
\left(\frac{f}{f_d(L)}-1\right)^2
-\dfrac{\pi c}{6}\dfrac{1}{L},
& f>f_d(L),\\[10pt]
\displaystyle
-\dfrac{\pi c}{6}\dfrac{1}{L},
& f<f_d(L).
\end{cases}
\label{eq:FL_bending}
\end{equation}
The terms proportional to \(-\pi c/6\) represent the free energy of the non-condensed
sector, namely the contribution already contained in \(W_{\rm CFT}[0]\). They follow from modular invariance of the CFT partition function on the torus in the large-\(c\) limit (see \cite{Apolo2022, KrausSivaramakrishnanSnively2017}),
\begin{equation}
	Z_{\rm CFT}(\beta,L)\simeq
	\begin{cases}
	\displaystyle
	e^{\frac{\pi c}{6}\frac{\beta}{L}},
	& \beta>L,\\[8pt]
	\displaystyle
	e^{\frac{\pi c}{6}\frac{L}{\beta}},
	& \beta<L .
	\end{cases}
\end{equation}

The transition between the two condensed branches is determined by
\begin{equation}
	F_T(f,L,T)=F_L(f,L,T) .
    \label{bending}
\end{equation}
It is useful to write this condition in dimensionless form. Define
\begin{equation}
	y=2-2\Delta,
	\qquad
	s=LT,
	\qquad
	\eta=\frac{f}{f_d}
	=
	A(\Delta)|f|L^{2-2\Delta}.
\end{equation}
Then
\begin{equation}
	\frac{f}{f_c}
	=
	\eta\,s^{-y}.
\end{equation}
The equality of the two free energies becomes
\begin{equation}
	s^2
	\left[
	\frac{\pi c}{6}
	+
	\frac{3A(\Delta)^2}{2B_\beta(\Delta)}
	\left(\eta s^{-y}-1\right)^2
	\right]
	=
	\frac{\pi c}{6}
	+
	\frac{3A(\Delta)^2}{2B_L(\Delta)}
	\left(\eta-1\right)^2 .
	\label{eq:bending_dimensionless}
\end{equation}
This equation makes the role of the quartic coefficients transparent. If
\begin{equation}
	B_\beta(\Delta)=B_L(\Delta)
\end{equation}
then \(s=1\), namely $T_{\rm SD}=\frac{1}{L}$,
is always a solution of \eqref{eq:bending_dimensionless} for any value of \(f\) in the
condensed phase. This is the expected result in the perfectly modular case.

If instead \(B_\beta(\Delta)\neq B_L(\Delta)\), the solution \(s=1\) is generically shifted.
To see this explicitly, set
\begin{equation}
	K_\beta=\frac{3A(\Delta)^2}{2B_\beta(\Delta)},
	\qquad
	K_L=\frac{3A(\Delta)^2}{2B_L(\Delta)},
	\qquad
	C=\frac{\pi c}{6}.
\end{equation}
Near the superconducting endpoint it is useful to set
\begin{equation}
	\eta=1+\delta_f,
	\qquad
	|\delta_f|\ll 1,
\end{equation}
and to look for a solution close to the undeformed self-dual point,
\begin{equation}
	s=1+\varepsilon,
	\qquad
	|\varepsilon|\ll 1.
\end{equation}
Expanding \eqref{eq:bending_dimensionless} to the first non-trivial order in
\(\delta_f\), one finds
\begin{equation}
	\varepsilon
	\simeq
	\frac{K_L-K_\beta}{2C}\,\delta_f^2
	+\mathcal O(\delta_f^3).
\end{equation}
Equivalently,
\begin{equation}
	T_{\rm SD}(f)
	\simeq
	\frac{1}{L}
	\left[
	1+
	\frac{9A(\Delta)^2}{2\pi c}
	\left(
	\frac{1}{B_L(\Delta)}-\frac{1}{B_\beta(\Delta)}
	\right)
	\left(\frac{f}{f_d}-1\right)^2
	\right].
	\label{eq:TSD_near_endpoint_explicit}
\end{equation}
This expression shows explicitly that the bending starts only at quadratic order in the
distance from the superconducting critical point. This is natural: near the transition the
condensate contribution to the free energy is already quadratic in the distance from
criticality, while the normal Hawking--Page free-energy difference varies linearly with
\(LT-1\).\\
Thus, while in the normal phase the self-dual transition is fixed by modularity, in the
condensed phase the branchwise GL description allows for a displacement of the transition
temperature if the quartic coefficients in the two asymptotic regimes are different.

This conclusion should nevertheless be interpreted with care, since it is obtained by
comparing the free energies in the two asymptotic regimes \(\beta\ll L\) and
\(\beta\gg L\), and assuming that these approximations remain reliable up to their crossing
point, as discussed in Subsection \ref{subsec:range_validity}. For this reason, even at large \(c\), the resulting curve should be regarded as an effective branchwise estimate of the
self-dual transition rather than as an exact finite-\(c\) phase boundary.

\subsection{Comparison with the AdS side}
\label{subsec:ads_comparison}

We can now compare the field theory model with its holographic counterpart.  
The first ingredient is the identification between the double-trace coupling $f$ introduced in the CFT effective description and the holographic boundary coupling $\kappa$. Matching the critical temperature obtained from the field theory model \eqref{tc toy model} with the one found on the AdS side \eqref{tc ads}, and identifying the CFT scaling dimension $\Delta$ with the AdS dimension $\Delta=\Delta_-$, gives (using the identity $\Gamma(x)=(x-1) \, \Gamma(x-1)$)
\begin{equation}
	\label{eq:f_kappa_dictionary_article}
	f=\frac{1-\Delta}{\pi}\,\kappa
	=\frac{\sqrt{1+m^2}}{\pi}\,\kappa .
\end{equation}
In the holographically renormalized normalization of the one-point function, it is convenient to use the coupling $f'=f/(2-2\Delta)$ (see, for example, \cite{Skenderis:2002wp, multitrace}); in that normalization the dictionary becomes $\kappa=2\pi f'$. For the present discussion, however, it is more convenient to keep the notation in terms of $f$.

From this identification it also follows naturally that the quantum critical point $f_d$ in \eqref{fd toy model} on the field theory side is identified with the critical coefficient $\kappa_v$ \eqref{eq kappa v} beyond which the condensate develops on the gravitational side.

Let us first compare the order parameter $\langle\mathcal{O}\rangle$.   
Near the transition, the field theory model predicts the standard mean-field behaviours
\eqref{behaviour alta T} and \eqref{behaviour bassa T}. In the form used below, both
branches are written as coupling-driven instabilities. In the high-temperature branch one
fixes the temperature and varies the double-trace coupling, so that the condensate is
controlled by the distance from the critical coupling $f_c(T)$. Through the dictionary
\eqref{eq:f_kappa_dictionary_article}, this is equivalently the distance from the
corresponding holographic critical coupling $\kappa_c(T)$ (obtained by reversing \eqref{tc ads}). In the low-temperature branch
the same logic applies with the critical value $f_d$, or equivalently $\kappa_v$ \eqref{eq kappa v}.

Thus the high-temperature condensed branch should be compared with the hairy BTZ
solutions, while the low-temperature condensed branch should be compared with the hairy
thermal AdS/solitonic solutions.

On the AdS side (see \cite{bolognesi}), the condensate is obtained numerically by solving the full backreacted
equations of motion, since the probe geometry is reliable only sufficiently close to the
critical point. More precisely, one solves the coupled bulk equations with $G_N\neq 0$ and
then extracts the expectation value from the near-boundary behaviour of the scalar profile
$H(\psi)$ through
\begin{equation}
	\alpha_H=
	\lim_{\psi\to \frac{\pi}{2}}
	\frac{(\tan\psi)^{1-2\sqrt{1+m^2}}}{2\sqrt{1+m^2}}\,
	\Big[H(\psi)(\tan\psi)^{\Delta_+}\Big]' \cos^2\psi .
\end{equation}
In the comparison shown below we use the same parameters as in the holographic analysis,
\begin{equation}
	m^2=-0.9,
	\qquad
	G_N=0.1.
\end{equation}
For the solitonic branch one fixes $\kappa_v\simeq -0.40775$ and varies $\kappa$, while
for the black hole branch one fixes $T$ and varies $\kappa$ across the critical value
$\kappa_c(T)$. In both cases the dominant uncertainty comes from the extraction of the scalar tail near the AdS boundary, estimated by varying the fitting windows and combining it with the statistical error of the tail fit. This uncertainty is propagated to the fitted condensate amplitude and hence to the quartic coefficient $B(\Delta)$.

\begin{figure}[p!]
	\centering
    \includegraphics[width=0.9\linewidth]{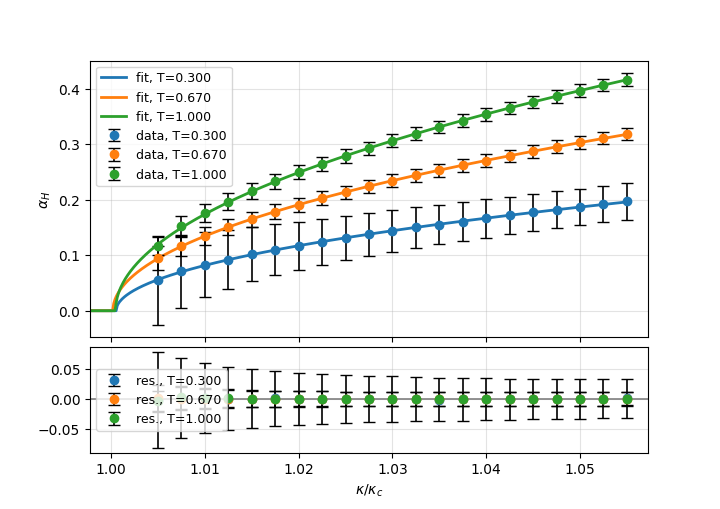}
    \includegraphics[width=0.9\linewidth]{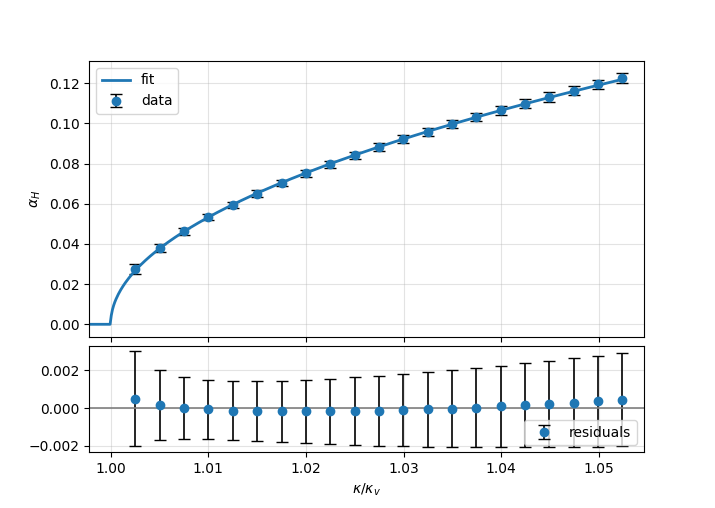}
    \caption{Comparison between the holographic condensate $\alpha_H$ and the field theory prediction $\langle\mathcal{O}\rangle$ near criticality. Top: BTZ-with-hair branch at different fixed temperatures, fitted with the high-temperature expression \eqref{behaviour alta T} as a function of the distance from $\kappa_c(T)$. Bottom: low-temperature solitonic branch, fitted with \eqref{behaviour bassa T} as a function of the distance from $\kappa_v$. The lower panels show the corresponding residuals.}
	\label{fig:condensate_comparison}
\end{figure}

Figure \ref{fig:condensate_comparison} shows the comparison with the holographic data.  
In both branches the square-root scaling predicted by the effective model is reproduced close to the corresponding critical point: $\kappa_c(T)$ in the BTZ branch and $\kappa_v$ in the solitonic branch. The agreement, however, is genuinely local. Already at moderately larger distances from the critical coupling the fitted curve starts to deviate from the numerical data, even though the condensate is still relatively small on the scale of the full solution. This is a useful indication of the limited regime of validity of the Ginzburg--Landau approximation used here.
This behaviour is expected. The field theory model keeps only the leading near-critical terms in the effective potential, while away from the transition higher powers of the order parameter and non-universal contributions become relevant. On the AdS side the same regime corresponds to the onset of nonlinear backreaction, which is not captured by the simple analytic form of \eqref{behaviour alta T} and \eqref{behaviour bassa T}. Thus the comparison should be interpreted as a test of the critical behaviour and of the near-critical amplitudes, not as a global fit of the full condensate curve.

The normalization of the condensate depends on the quartic coefficient appearing in the
Ginzburg--Landau effective action. In the branchwise approximation used here, the two
asymptotic regimes give two effective values,
\begin{equation}
	B_\beta(\Delta)=837 \pm 20,
	\qquad
	B_L(\Delta)=742 \pm 12 .
\end{equation}
These numbers are extracted from the holographic condensate fits in the BTZ and solitonic
branches, respectively.

The fact that the two values are not equal should not be interpreted as a breakdown of
modularity. As discussed in Subsection \ref{subsec:landau_condensate}, modular covariance is
expected to be a property of the complete torus partition function, not of each isolated
saddle in a truncated near-critical expansion. In the present comparison we are instead
treating separately the two extreme semiclassical branches. On the AdS side, the
backreaction of the scalar condensate acts differently on the BTZ and solitonic geometries;
on the field theory side this difference is naturally encoded in the non-universal quartic
coefficients $B_\beta(\Delta)$ and $B_L(\Delta)$. Thus the mismatch between the two
effective values reflects the branchwise nature of the approximation, rather than a failure
of modularity in the full theory. 

\begin{figure}[p!]
    \centering 
    \includegraphics[width=1\linewidth,height=0.65\textheight]{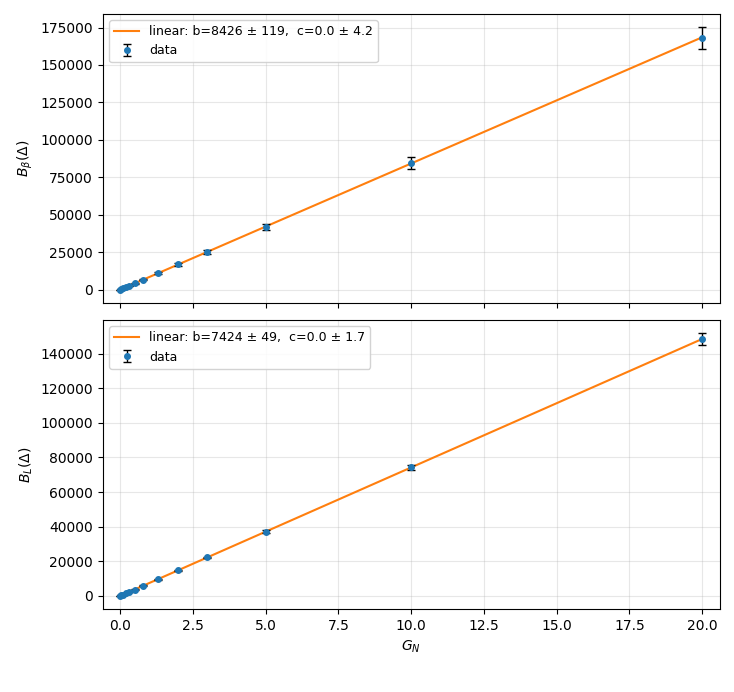}
	\caption{Quartic coefficients extracted from the holographic condensate as functions of $G_N$. The BTZ coefficient $B_\beta(\Delta)$ is obtained at fixed temperature $T=1$, while $B_L(\Delta)$ is obtained from the solitonic branch. The data are fitted with $B_i(\Delta)=b_iG_N+c_i$.}
	\label{fig:GN_dependence}
\end{figure}

The bulk analysis also makes it possible to study directly how the quartic coefficients depend on the backreaction parameter $G_N$. Repeating the numerical extraction of the condensate for several values of $G_N$, we obtain the coefficients $B_\beta(\Delta)$ and $B_L(\Delta)$. For the BTZ branch the extraction is performed at fixed temperature $T=1$. As a preliminary check, we fitted the data with
\begin{equation}
	B_i(\Delta)=b_i\,G_N^{\, p_i}+c_i,
	\qquad i=\beta,L .
\end{equation}
The resulting exponents are compatible with $p_i=1$. Therefore, in Figure \ref{fig:GN_dependence} we fix the exponent to one and fit the data with the linear law
\begin{equation}
	B_i(\Delta)=b_i\,G_N+c_i,
	\qquad i=\beta,L .
\end{equation}
The linear dependence and the fitted intercepts compatible with zero support the interpretation that the leading bulk gravitational effect is encoded in the quartic coefficient of the effective Ginzburg--Landau theory. This is consistent with the field theory analysis, where $B_i(\Delta)$ is sensitive to finite-$c$ corrections, which are absent in the strict large-$c$ limit. Through the Brown--Henneaux relation $c=3/(2G_N)$ \cite{BrownHenneaux1986}, which for example gives $c=15$ when $G_N=0.1$, this is equivalently the leading dependence on $G_N$.\\
From the bulk point of view, this interpretation is natural. Around the normal background, where the scalar field is small, the scalar stress tensor starts at order $|\Phi|^2$, and the Einstein equations therefore generate metric corrections of order $G_N|\Phi|^2$. Substituting the corrected solution into the on-shell action then produces an effective quartic contribution controlled by $G_N$. In the zero-current sector considered here, the gauge field is trivial and does not provide an independent contribution to this coefficient.\\
It is important not to overinterpret the large-$G_N$ region. By the Brown--Henneaux relation, large $G_N$ corresponds to small central charge, where both the semiclassical bulk description and the large-$c$ factorization assumptions are outside their controlled regime. Thus the persistence of the linear trend at large $G_N$ should be regarded as a feature of the classical model and of the near-critical fit, while the physical interpretation of the scaling is reliable only at small $G_N$, or equivalently large $c$.

We now turn to the comparison of the phase diagram. 
With the dictionary \eqref{eq:f_kappa_dictionary_article}, the two descriptions admit a
natural phase-by-phase identification. The low-temperature branch of the toy model, namely
the regime dominated by the spatial circle, corresponds to the thermal AdS/solitonic sector,
while the high-temperature branch corresponds to the BTZ sector. The four phases are
therefore mapped as follows:
thermal AdS without hair $\leftrightarrow$ low-$T$ phase without condensate,
thermal AdS with hair $\leftrightarrow$ low-$T$ phase with condensate,
BTZ without hair $\leftrightarrow$ high-$T$ phase without condensate,
and BTZ with hair $\leftrightarrow$ high-$T$ phase with condensate.

\begin{figure}[t!]
	\centering
    \includegraphics[width=0.9\linewidth]{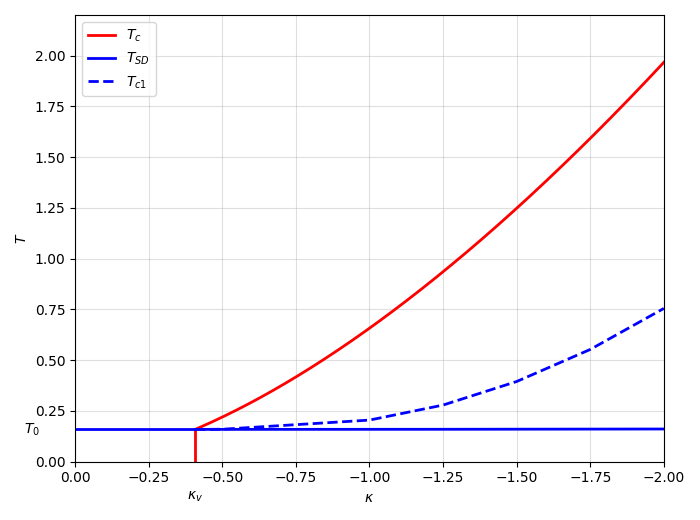}
	\caption{Phase diagram of the field theory model in the $( \kappa ,T )$ plane after using
	the dictionary \eqref{eq:f_kappa_dictionary_article}. For comparison with the AdS side,
	the dashed curve represents the critical line $T_{c1}$ shown in Figure
	\ref{diagramma di fase}.}
	\label{fig:phase_diagram_comparison}
\end{figure}

Figure \ref{fig:phase_diagram_comparison} summarizes the resulting picture and should be compared with the holographic phase diagram in Figure \ref{diagramma di fase}. In the normal phase, the self-dual transition starts at $T_{\rm SD}=1/(2\pi)$, exactly as in the Hawking--Page transition of the AdS model. Inside the condensed phase, the field theory model predicts a bending of the self-dual transition \eqref{eq:TSD_near_endpoint_explicit}, due to the condensate contribution to the free energy in the two modular regimes.
In the present numerical implementation, however, this effect is confined to a very small interval of $\kappa$ close to $\kappa_v$. On the scale of the full phase diagram the curve is therefore almost horizontal. Moreover, this near-critical region is precisely the one where no detailed holographic data for the corresponding line are available, so the comparison with the dashed curve $T_{c1}$ should not be interpreted quantitatively. Away from $\kappa_v$, where the AdS curve develops its visible bending, the Ginzburg--Landau expansion is no longer expected to be reliable: the condensate is no longer small and the transition is controlled by the full nonlinear bulk solution. Thus the field theory result should be understood as a local near-critical prediction for the onset of the bending, rather than as a reconstruction of the complete holographic curve.

By contrast, the superconducting transition obtained from the quadratic instability is more
robust in the asymptotic regimes. Strictly speaking, on a generic torus the full transition
line is not known analytically; however, under the assumptions stated in Subsection
\ref{subsec:range_validity}, the high- and low-temperature branches provide the correct
instability line on the two sides of the self-dual point.

To summarize, once the relation between $f$ and $\kappa$ is fixed, the field theory model
reproduces the same qualitative phase structure as the AdS analysis and captures with good
quantitative accuracy the behaviour of the condensate near the critical lines. The comparison
also shows that the quartic Ginzburg--Landau coefficients encode the leading sensitivity to
gravitational backreaction, with distinct effective values in the BTZ and solitonic branches.
This branch dependence should be understood as a feature of the near-critical,
saddle-by-saddle approximation, not as a breaking of modularity of the full theory. The main
limitation is also clear: the CFT description is designed to be reliable in the critical region,
so its predictions should not be extrapolated too far from the transition. Within this regime
of validity, the comparison supports the interpretation of the model as an effective boundary
description of the AdS$_3$ holographic superconductor.

\subsection{Comments on the condensate away from criticality}
\label{subsec:condensate_away}

The comparison with the holographic condensate shows that the square-root behaviour is
accurate only very close to the critical line. This is not surprising, since the expressions
used above come from a quartic truncation of the near-critical effective action. It is useful,
however, to check how the same expansion looks in a different but equivalent
parametrization of the effective theory.

In the discussion above we expanded the connected functional \(W_{\rm CFT}[\sigma]\) and
then used the saddle relation \(\langle\mathcal O\rangle=-\sigma/f\). Equivalently, one may
work with the 1PI effective action for the order parameter
\(\varphi=\langle\mathcal O\rangle\). Keeping only the homogeneous mode, let
\begin{equation}
	W_{\rm CFT}[\sigma]
	=
	\frac{1}{2}G_0\,\sigma^2
	-
	\frac{u}{4!}\,\sigma^4+\cdots .
\end{equation}
The Legendre transform gives, to quartic order,
\begin{equation}
	\Gamma_0[\varphi]
	=
	\frac{1}{2G_0}\,\varphi^2
	+
	\frac{u}{4!\,G_0^4}\,\varphi^4+\cdots .
\end{equation}
The factors of \(G_0^{-1}\) have the usual interpretation of amputated external legs:
the connected four-point coefficient appearing in \(W_{\rm CFT}[\sigma]\) is converted into the
corresponding 1PI vertex by removing the four external two-point functions.

After adding the double-trace deformation, the quadratic coefficient in $\Gamma_f[\varphi]$ becomes
\(G_0^{-1}+f\). Minimizing the 1PI potential therefore gives
\begin{equation}
	\varphi_{\rm 1PI}^2
	=
	-\frac{6G_0^4}{u}\left(G_0^{-1}+f\right).
\end{equation}
For example on the low-temperature branch, where \(G_0=1/|f_d|\), this differs from the result obtained from the \(\sigma\)-effective action \eqref{eq:condensate_general} by the factor
\begin{equation}
	\langle\mathcal O\rangle_{\rm 1PI}^2
	=
	\left(\frac{f}{f_d}\right)^3
	\langle\mathcal O\rangle_{\sigma}^2 .
\end{equation}
Equivalently, the condensate differs by a factor \((f/f_d)^{3/2}\). This factor is precisely
what is lost when, in the near-critical expression obtained from \(W_{\rm CFT}[\sigma]\) \eqref{behaviour bassa T}, one
replaces the running coupling in the prefactor by its critical value, \(f\simeq f_d\). With
this standard near-critical approximation the two parametrizations give the same leading
condensate. The same statement holds in the high-temperature branch, with \(f_d\) replaced
by the temperature-dependent critical coupling \(f_c(T)\). This does not mean that one description is more fundamental than the other. The full
functionals \(W\) and \(\Gamma\) are related by a Legendre transform and contain the same
information. The difference appears because the expansion has been truncated at quartic
order, and the Legendre transform does not commute with this truncation. Close to the
transition the factor is subleading, while farther away it changes the shape of the curve.

This alternative parametrization does not improve the comparison with the AdS side in any essential way. The main point is that the deviation from the holographic condensate is not simply a matter of choosing
between \(W\) and \(\Gamma\). Once the condensate is no longer small, higher powers
of the order parameter and, especially, non-universal bulk data, become important. This is also visible from the self-dual line. If instead of the strict Ginzburg--Landau square-root profile one
uses the actual condensate behaviour extracted from a free fit to the holographic data, the
resulting bent self-dual curve becomes much closer to the AdS curve. This confirms that the
mismatch is mainly due to the limited validity of the near-critical condensate profile, not to
the general structure of the effective description. 

The absence of a simple universal continuation away from the transition should be
contrasted with the mechanism discussed in \cite{multitrace}. There, when the critical
temperature is driven to zero at finite density, the condensate can be controlled by matching
the UV double-trace boundary condition to an emergent IR region, typically AdS$_2$, and
the scaling is fixed by the dimension of the scalar operator in the corresponding IR CFT.
In the present AdS$_3$ model there is no analogous regular IR conformal region controlling
the whole crossover. The solitonic branch caps off in the interior, while the \(T\to0\)
limit of the hairy black hole is a zero-entropy configuration with logarithmic behaviour in
the geometry \cite{bolognesi}. Thus one should not expect a simple universal scaling form
for the condensate far from the transition. The robust statement is the near-critical one:
the effective CFT description captures the mean-field exponent and the local amplitude, but
not the full nonlinear condensate curve.

\section{Field Theory toy model for the fractional Little-Parks effect}
\label{sec:LP}

In this section we consider a toy model that can reproduce some of the  features observed in the holographic model  regarding vortices and the Little-Parks effect. We are interested in reproducing in a simple setup the non-integer value of the magnetic flux, which corresponds to the holonomy of the gauge field in the $1+1$-dimensional theory.  

We take a $1+1$-dimensional $U(1)$ gauge theory with two charged fields, both have a symmetry-breaking potential whose minimum is non-zero
\beq
S_{1+1} = \int d^2 x \left( -\frac{1}{4 g^2} F^2 + |D\varphi|^2 - \lambda^2 (|\varphi|^2-v^2)^2 + |D\Phi|^2 - \Lambda^2 (|\Phi|^2-V^2)^2 
\right) 
\eeq 
where $D = \partial - i A$, we take for simplicity  the same charge for the two fields.
The model has two $U(1)$ symmetries. One is gauged and corresponds to the electromagnetic symmetry
\beq
U(1)_e : \ \varphi \to e^{i \alpha} \varphi \quad \Phi \to  e^{i\alpha} \Phi \ ,
\eeq
the other one is global and is  
\beq
U(1)_a : \ \varphi \to e^{i \beta} \varphi \quad \Phi \to  e^{-i\beta} \Phi \ .
\eeq
$U(1)_a$ could be broken explicitly if we add a direct interaction $V_{\rm int}$ between the two fields, but we do not do it for the moment, and  we set $V_{\rm int}=0$.  The parameters of the model are $g$, $\lambda$, $\Lambda$ (with dimension of a mass) and $v$, $V$ (dimensionless). 
We consider energy scales  $E \ll \Lambda V $ so that $\Phi = V e^{i \theta}$ is fixed at the minimum of its potential.   
We also consider energy scales  $E \ll g$ so the effect of $A_{\mu}$ is to remove a phase degree of freedom and acts like a Lagrange multiplier.
So we have a reduced model
\beq
S_{1+1} = \int d^2 x \left(   |D\varphi|^2 - \lambda^2 (|\varphi|^2-v^2)^2 + |D\Phi|^2 \right)
\label{S1+1}
\qquad 
\label{Vinfty}
\eeq 
with the constraint $|\Phi|^2 =V^2$.
There is a Higgs field, the fluctuation of $|\varphi|$ around the minimum $v$, with mass  $2 \lambda v$.
We have one massless scalar field as the gauge constraint removes one degree of freedom.

We consider the theory on $R \times S^1$, we identify $x \simeq x + 2 \pi R$. 
We turn on a winding for the two fields \beq\varphi = w e^{\frac{i n x}{R}}, \qquad \Phi = e^{\frac{i m x}{R}} V, \eeq 
where $(n,m) \in \mathbb{Z}^2$ are both integers. We can set  $m=0$ and vary $n$.   The windings   induce a Wilson line for the gauge field. In the gauge $A_t=0$, we set  $ A_x = a $, with $a$ constant. We consider a translationally invariant ansatz with $w$ and $a$.
The energy is 
\beq
E[w,a] = 2 \pi R \left( \left(\frac{n}{R} - a\right)^2 w^2 + \lambda^2 (w^2-v^2)^2 + a^2 V^2\right).
\label{Eaw}
\eeq
Minimizing with respect to $a$, we obtain
\beq
a(w) = \frac{n w^2}{R(V^2 + w^2)} 
\eeq
then we have
\beq
E[w] = 2 \pi R \left( \left(\frac{n}{R} - a(w)\right)^2 w^2 + \lambda^2 (w^2-v^2)^2 + a(w)^2 V^2\right)
\eeq
to minimize for $w$.
The Wilson line and the scalar condensate are always in the range
\beq
0 \leq a \leq  \frac{n  }{R }, \qquad \quad v \geq w \geq 0,
\eeq
for $n=0$,  $a=0$ and $w=v$.   Increasing $n$ reduces the condensate and there is a maximum $n$ after which $w=0$, $a=0$ and $E=0$ and  superconductivity is lost.   
In the absence of the second field $\Phi$, or for $V = 0$, the Wilson line is exactly the one that cancels the covariant derivative $ a = \frac{n  }{R } $, and $w =v$, $E=0$.  
The solution for large $\lambda$ ($\lambda R \gg v$), and small $n$, is 
\beq
a \simeq \frac{n v^2}{R(V^2 + v^2)}, \qquad w \simeq v - \frac{n^2 V^2}{4 R^2 v (v^2 + V^2)\lambda^2}, \qquad 
E \simeq \frac{2 \pi n^2 v^2 V^2}{R (v^2 + V^2)}.
\eeq  
The energy scales with $n^2$, at least for $n$ small enough so that $v -w \ll v$.

There are two important parameters in this toy model, much like the holographic counterpart. One is the ratio $\frac{v}{V}$. This controls the value of 
\beq
\frac{a R}{n} =  \frac{\frac{v^2}{V^2}}{1+\frac{v^2}{V^2}} \qquad {\rm for } \ n \ {\rm small}.
\eeq 
In holography the parameter that controls the magnitude of the gauge field at infinity $ \propto e \alpha_H $. 
The other parameter of the toy model is $\lambda R v$
\beq
\frac{w}{v} \simeq 1 - \frac{n^2 }{4 (\lambda R v)^2 \left(1+\frac{v^2}{V^2}\right)} \qquad {\rm for } \ n \ {\rm small}.
\eeq
In the holographic counterpart this is $\frac{\alpha_H|_{n}}{\alpha_H|_{n=0}}$ and is related to both $\kappa$ and the gravity backreaction.

\begin{figure}[h!]
\centering
\includegraphics[width=0.5\textwidth]{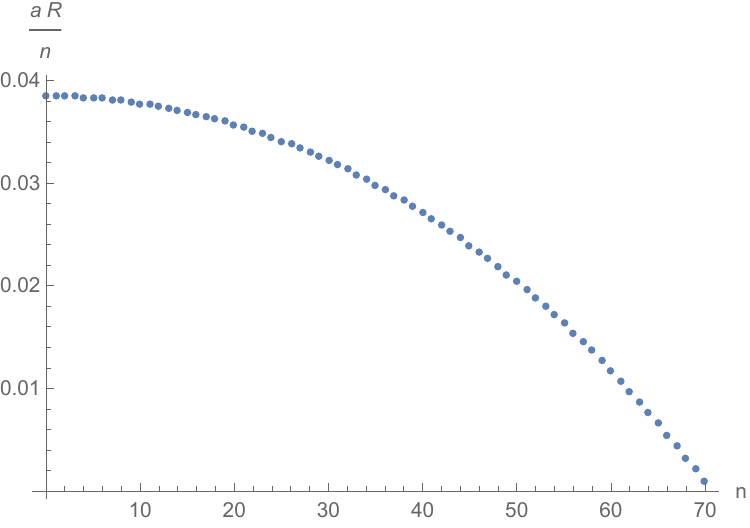}\quad
\includegraphics[width=0.5\textwidth]{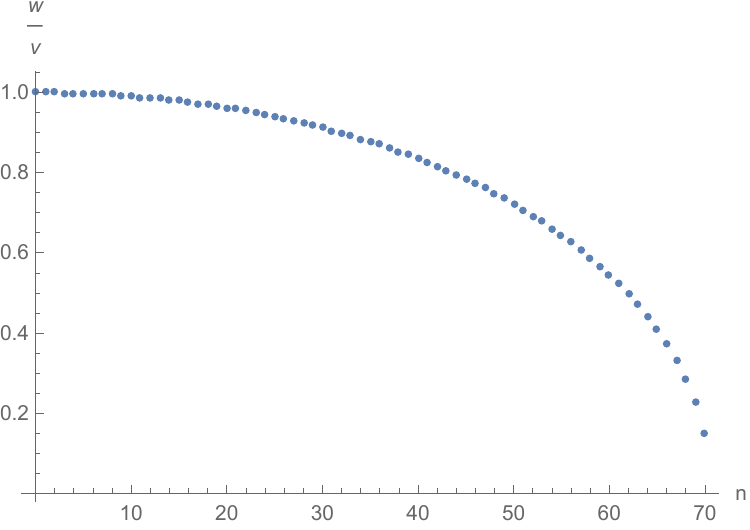}\quad
\includegraphics[width=0.5\textwidth]{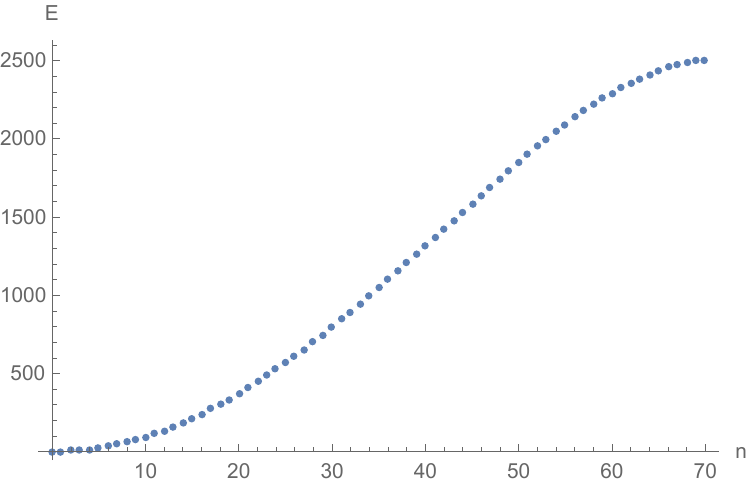}
\caption{  \small Plots of $\frac{a R}{n}$, $\frac{w}{v}$ and $E$ for  $v=1$, $V=10$, $R=1$  and $\lambda =50$.}
\label{vVlambda}
\end{figure}

We give some numerical solutions of the minimization problem in Figure \ref{vVlambda}. We take for reference $v=1$, $V=10$, $R=1$ and   $\lambda =50$  (in this case $n_{max} = 70$). Increasing $\lambda$ increases the value of $n_{max}$. The condensate $w$ goes to $0$ at $n_{max}$. These are also  features we encountered in the holographic model changing the double-trace deformation $\kappa$. For finite $\lambda$ $n$ is not a topological number, these solutions are in fact metastable and $n_{max}$ is when they do not exist anymore. As $\lambda \to \infty$, $n$ becomes a conserved topological number and $n_{max} \to \infty$.

In a physical setup, to realize the Little-Parks effect, we  think of this $1+1$-dimensional theory  as a circular superconducting wire in $2+1$-dimensional bulk (or a  superconducting hollow cylinder immersed in $3+1$ dimensions). For practical realization of a $1+1$-dimensional superconductor  there must be a  finite thickness of the wire in order to evade the CMWH theorem. Here  we neglect the thickness.
\begin{figure}[h!]
	\centering
	\includegraphics[width=0.5\textwidth]
    {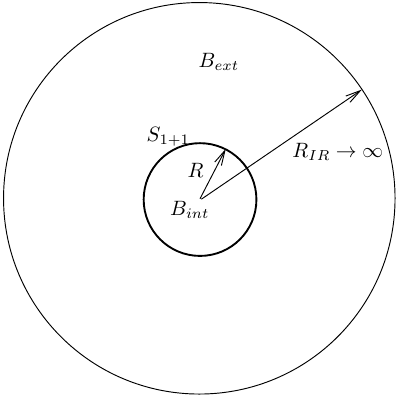}
 	\caption{  \small  Geometry of the circular wire of radius $R$ in an external magnetic field. The IR radius $R_{\rm IR}$ is used to set the boundary conditions for the external magnetic field.  }
	\label{figLP}
\end{figure} 
To realize the  Little-Parks effect we  need to add the bulk electromagnetic interaction and then put the system in an external magnetic field.  
A  circular superconducting wire with action $S_{1+1}$ is placed in a $2+1$-dimensional external electromagnetic field
and bulk$+$wire are described by the effective action
\beq
S = \int d^3 x \, \left(-\frac{1}{4 e^2} F^2  \right) + S_{1+1}\left[\varphi,\Phi,A_{\mu}\right]
\eeq 
where $S_{1+1}$ for us is \eqref{S1+1}.
The electromagnetism can propagate in the $2+1$-dimensional bulk space-time outside the wire and interact with the  $1+1$-dimensional theory.  We take a radially symmetric ansatz in polar coordinates with a field $A_{r}=0$, $A_{\theta}(r)$ generic and magnetic field $B = \frac{1}{r} F_{r\theta} $,  $F_{r\theta} = \partial_r A_{\theta}$. For the wire we take homogeneous ansatz with $\varphi = w$ and $A_{\theta}(R) = a R $, as before.   We consider an IR cutoff $R_{\rm IR} \gg R$ with boundary condition $A_{\theta} (R_{\rm IR}) = B_{\rm ext} R_{\rm IR}^2$ and $A_{\theta}(0) = 0 $. Taking $R_{\rm IR} \to \infty$ is equivalent to placing the wire in an external magnetic field $B_{\rm ext}$. 
Without the wire, or with the wire in some special optimal conditions, magnetic field is constant, $B_{\rm int} = B_{\rm ext}$ and  $A_{\theta} (r) = B_{\rm ext} r^2$ everywhere.  
Solving the bulk equations we have in general
\beq
 A_{\theta} = \left\{\begin{array}{lll}
	 a \frac{r^2}{R},   & &r \in (0,R)   \\   \alpha r^2 + \beta,  &   &r \in (R,R_{\rm IR})
\end{array}\right.
\eeq
with 
\beq
\alpha =  \frac{B_{\rm ext} - a \frac{R}{R_{\rm IR}^2}}{1-\frac{R^2}{R_{\rm IR}^2}}, \qquad \beta = a R - \alpha R^2
\eeq
where we imposed that the magnetic field outside is $B_{\rm ext}$ and $A_{\theta}$ is continuous on the wire.
 There may be a discontinuity in the derivative of $A_{\theta}$ in $R$; this can produce a difference between the internal and external magnetic field. 
The total energy, including only the leading terms for  $R_{\rm IR} \gg R$, is
\beq
E = \pi B_{\rm ext} R_{\rm IR}^2 + a^2 \pi - 2 \pi B_{\rm ext} a R  
+  E_{1+1}[\varphi,\Phi,A_{\mu}]
\eeq   
with
\beq
2 \pi (a - R   B_{\rm ext})   = - \frac{d E_{1+1}}{d a} .
\eeq
The difference between the external and internal magnetic field $B_{\rm int}- B_{\rm ext}$ is the direct  source for the gauge field $a$, analogous to $J$ in the holographic model. 
In the standard physical realization of the LP effect we keep  $B_{\rm ext}$ as the external control parameter and see how the system responds to its variation. In the holographic problem instead we   keep fixed $ B_{\rm ext} - B_{\rm int} $. 
In 'optimal conditions'  
\beq
B_{\rm ext} = B_{\rm int}, \qquad a = B_{\rm ext} R, \qquad \frac{d E_{1+1}}{d a} = 0.
\eeq 
This is the case where we have zero source for the $1+1$-dimensional theory and correspond to the stationary solutions of $S_{1+1}$ in isolation, discussed before.

\begin{figure}[h!]
	\centering
	\includegraphics[width=0.5\textwidth]{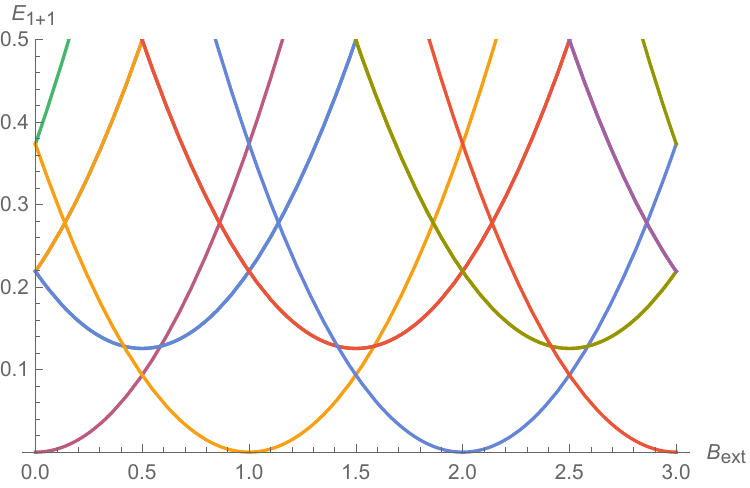}\qquad
	\includegraphics[width=0.5\textwidth]{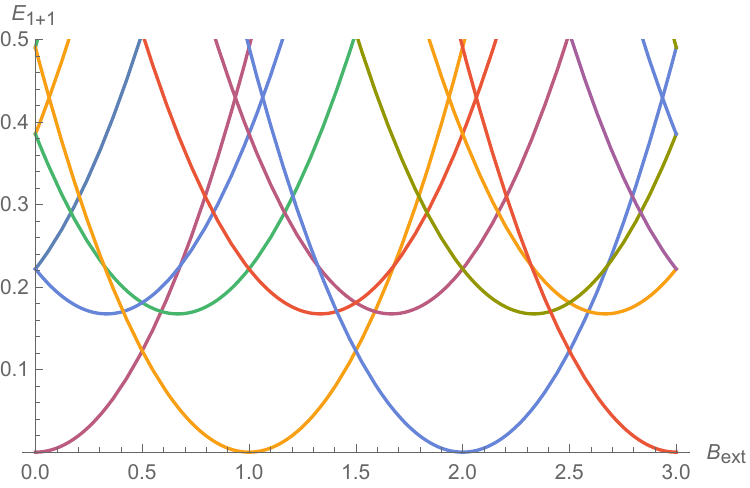}
	\caption{  \small Plots of $ E_{1+1}$ 
for $v=0.2$ and $V=0.2$ (left) or $V= 0.2 \  \sqrt{2} $ (right). In this case apart from the usual LP periodicity there are states with $1/2$  flux (left) or $1/3$  flux (right). }
	\label{figLP2}
\end{figure}

We consider for simplicity  just the $\lambda \to \infty$ limit where $w =v$
\beq
E = \pi B_{\rm ext} R_{\rm IR}^2  + a^2 \pi - 2 \pi B a R  + E_{1+1}
\eeq
where
\beq
E_{1+1} =  2 \pi R \left( \left(\frac{n}{R} - a\right)^2 v^2 +  \left(\frac{m}{R} - a\right)^2  V^2\right).
\label{Eaw2}
\eeq
Now we just have to minimize to find 
\beq
a= \frac{B_{\rm ext} R+2 m V^2+2 n v^2}{2 R v^2+2 R V^2+1}, \qquad B_{\rm int} = \frac{a}{R},
\eeq
and the solution for optimal conditions  $B_{\rm ext}=B_{\rm int} = \frac{a}{R}$  is $
a  =\frac{m V^2+n v^2}{R \left(v^2+V^2\right)} $.
For $n=m$ we have the case $a = \frac{n}{R} $, and $  E_{1+1} =0$. This  corresponds to the usual periodicity of the LP effect, and thus to the hairy black hole state in holography. For other choices of $n-m \neq 0$ we can have fractional magnetic flux and this corresponds to the solitonic vortex in holography. Simple illustrative cases, when $v=V$ or $V=\sqrt{2}v$, are given in Figure \ref{figLP2} where the fractional fluxes are $1/2$ or $1/3$. 
The energy $E_{1+1}$ at zero $B_{\rm int} - B_{\rm ext}$ is always minimized by the integer flux states. This remains true even changing $\frac{v}{V}$.

The Little-Park periodicity corresponds to $(n,m) \to (n+k,m+k)$ both translated by the same integer $k \in \mathbb{Z}$.
For solitons, in absence of a Dirac string $a(0)=0$, 
the symmetry of the equations of motion $n\to n+p, \ a(r) \to a(r)+p$ 
is broken explicitly by  the boundary conditions in eq.(\ref{boundarysoliton}).
 If we allow for the presence of a Dirac string singularity $a(0)$
 in \eqref{boundarysoliton}, we can also realize the LP periodicity  for solitons in AdS.
In holography we have a similar situation in the soliton phase.  The holographic winding of the soliton corresponds to the difference $m-n$. So $m=n$ is the ground state of the soliton and the periodicity of LP is realized by the insertion of the Dirac string in the bulk. The other states with $n-m \neq 0$ correlate with the vortices with fractional flux.

\section{Conclusions}
\label{sec:conclusion}

In this paper we studied 
two field theory models motivated by the AdS$_3$/CFT$_2$
holographic superconductor of \cite{bolognesi}. The first one addresses the homogeneous
sector of the theory, namely the phase diagram and the onset of the condensate. The second
one focuses instead on the vortex sector and provides a field theory mechanism for
the fractional-flux effect observed in the solitonic phase.

The main result of the first model is that several features of the holographic phase diagram
can be recovered from a large-$c$ CFT deformed by a relevant double-trace operator. The
condensation is interpreted as an instability of the quadratic kernel of the effective action.
In the high- and low-temperature cylinder limits, this instability can be computed
analytically from the two-point function of the undeformed CFT, reproducing the structure
of the holographic critical line after the identification between the double-trace coupling
$f$ and the Robin boundary parameter $\kappa$. The comparison also clarifies the dictionary for the source and the VEV: with our Euclidean
conventions, \(\sigma\sim-\beta_H\) while \(\langle\mathcal O\rangle\sim\alpha_H\).

Moreover, modularity gives the field theory interpretation of the Hawking--Page transition. In the
normal phase, the exchange of the two torus cycles fixes the self-dual point, while the
superconducting instability line intersects it at the point where the three scales of the
problem meet: temperature, finite-size gap, and double-trace scale. Under the additional
assumptions discussed in Subsection \ref{subsec:range_validity}, the two asymptotic branches
can be extended up to this point, giving a field theory reconstruction of the four-phase structure of the AdS model.

The same effective description also captures the condensate near the transition. Including
the leading quartic term in the large-$c$ effective action gives a Ginzburg--Landau
description with the expected mean-field square-root behaviour. The comparison with the
backreacted AdS solutions shows good quantitative agreement close to the critical couplings
$\kappa_c(T)$ and $\kappa_v$. The comparison also identifies the quartic coefficients
$B_\beta(\Delta)$ and $B_L(\Delta)$ as the non-universal data controlling the amplitude of
the condensate in the two asymptotic branches.
These coefficients are not fixed by conformal symmetry alone, since they depend on
connected four-point data of the underlying CFT. On the holographic side they encode the
leading sensitivity to gravitational backreaction. This is supported by the numerical
extraction of $B_\beta(\Delta)$ and $B_L(\Delta)$ at different values of $G_N$, which shows
a linear dependence with intercept compatible with zero. This scaling should be interpreted
physically only in the small-$G_N$, large-$c$ regime; at large $G_N$ both the semiclassical
bulk description and the large-$c$ factorization assumptions are being extrapolated beyond
their controlled domain.

The limitations of the construction are equally important. The effective CFT model is
near-critical and branchwise. It correctly describes the onset of the instability and the local
behaviour of the condensate, but it should not be used as a global fit of the full nonlinear
condensate curve. This is also reflected in the bending of the self-dual line: the
Ginzburg--Landau analysis predicts the beginning of the bending close to the quadruple
point, but it does not reconstruct the full holographic curve away from criticality. In that
region higher powers of the order parameter and non-universal bulk data become important.

Beyond the specific holographic comparison, the first model identifies a general superconducting transition in two-dimensional CFTs with a large-$c$ limit.
Under the standard assumptions used throughout the analysis, the transition is driven by
the instability of a double-trace-deformed quadratic kernel, while the leading condensate
behaviour is controlled by a Ginzburg--Landau expansion. In the decompactification limit,
the finite-size critical coupling becomes a genuine double-trace quantum critical point, in
the same spirit as the holographic mechanism discussed in \cite{multitrace}.

The second model addresses a different feature of the AdS$_3$ construction: the existence
of vortex solutions with fractional magnetic flux in the solitonic phase. We introduced a weakly coupled $1+1$-dimensional gauge theory with two charged scalar fields. In
the regime where the gauge field is effectively auxiliary, the Wilson line is fixed dynamically
by minimizing the energy and need not take integer values. This gives a transparent field
theory mechanism for fractional flux, without interpreting it as a violation of the usual
quantization condition.

When the system is coupled to an external electromagnetic field, the integer-flux branches
reproduce the usual Little--Parks periodicity, while sectors with non-trivial relative winding
between the two charged fields give fractional-flux branches. In this sense the model
provides a fractional Little--Parks effect, which mirrors the distinction between the black
hole and solitonic vortex sectors in the holographic theory. The toy model also reproduces
qualitative features such as the suppression of the condensate as the winding increases and
the disappearance of vortex branches beyond a maximal winding.

At the same time, this second model is not meant to be a complete microscopic dual of the
holographic vortex sector. In particular, in the simple version studied here the energetically
preferred configurations at zero external source are still the integer-flux ones, whereas in
the holographic model solitonic vortices can dominate in part of the phase diagram. The
toy model should therefore be viewed as isolating the field theory mechanism behind
fractional flux, rather than reproducing the full saddle competition of the gravitational
system.

Several extensions would be natural. For the first model, the main open problem is to embed
the effective description into an explicit class of large-\(c\) CFTs, so that the quartic
coefficients \(B_\beta(\Delta)\) and \(B_L(\Delta)\) can be computed directly from boundary
correlators. Other directions include adding the gauge sector for a more direct comparison
with the AdS model, considering non-scalar double-trace perturbations, extending the
construction to higher dimensions, and going beyond the quartic Ginzburg--Landau
truncation to describe the nonlinear regime of the condensate. On the vortex side, one
could enrich the weakly coupled toy model with interactions or additional sectors capable
of reproducing not only fractional flux, but also the energetic dominance of solitonic vortex
states. Together, these directions would clarify how much of the AdS$_3$ holographic
superconductor can be understood from universal field theory principles, and which features
require genuinely model-dependent CFT data.

\section*{Acknowledgments}

We thank G. Nardelli and N. Zenoni for discussions and collaboration on related project.
The work  is supported by the INFN special research project
grant `GAST' (Gauge and String Theories).

\appendix

\section{Effective action in momentum space}
\label{app:momentum-space}

Using the Fourier series (with $V=\beta L$)
\begin{align}
	\sigma(x) &= \frac{1}{\sqrt V}\sum_{q} e^{\,i q\cdot x}\,\sigma_q,
	&
	\sigma_q &= \frac{1}{\sqrt V}\int_0^\beta d\tau \int_0^L dx\; e^{\,-i q\cdot x}\,\sigma(x),\\
	G_0(x) &= \frac{1}{V}\sum_{q} e^{\,i q\cdot x}\,G_0(q),
	&
	G_0(q) &= \int_0^\beta d\tau \int_0^L dx\; e^{\,-i q\cdot x}\,G_0(x),
\end{align}
together with the orthogonality relation
\begin{equation}
	\int_0^\beta d\tau \int_0^L dx\; e^{\,i(q+q')\cdot x}=V\,\delta_{q+q',0}\,,
\end{equation}
one obtains
\begin{align}
	&\int d^2x\, d^2y\ \sigma(x)\,G_0(x-y)\,\sigma(y)
	=  \frac{1}{V^2}\sum_{p}G_0(p)\sum_{q,q'}\!
	\int d^2x\, e^{\,i(q+p)\cdot x}
	\int d^2y\, e^{\,i(q'-p)\cdot y}\, \sigma_q\,\sigma_{q'} = \nonumber \\
	&= \frac{1}{V^2}\sum_{p}G_0(p)\sum_{q,q'}\!
	\big[V\delta_{q+p,0}\big]\big[V\delta_{q'-p,0}\big]\ \sigma_q\,\sigma_{q'} = \sum_{q}\sigma_{-q}\,G_0(q)\,\sigma_q
\end{align}
and also
\begin{equation}
	\int d^2x\, d^2y \ \sigma(x)\delta^{(2)}(x-y)\sigma(y) = \sum_{q}\sigma_{-q}\,\sigma_q \,.
\end{equation}

\section{Instability criterion for generic momenta \texorpdfstring{$m$}{m} and \texorpdfstring{$n$}{n}}
\label{m ed n generici}
For generic momenta, the full expression for \(J(\Delta,m,n)\) must be kept.
Starting from \eqref{eq:G0J} and modifying it according to the more general definition of the susceptibility in \eqref{suscettività}, one obtains
\begin{equation}
	J(\Delta,m,n)\equiv
	\int_{0}^{\pi} dv \int_{-\infty}^{\infty} du\;
	\frac{e^{2imu+2inv}}{\bigl(\sinh^2 u+\sin^2 v\bigr)^\Delta},
\end{equation}
and using the Schwinger representation as in the homogeneous case, the integrals \eqref{eq:vint} and \eqref{eq:uint} become
\begin{equation}
	\int_{0}^{\pi} dv\; e^{-t\sin^2 v}e^{2imv}
	=
	\pi e^{-t/2}\,\frac{1}{2}
	\left[
	I_m\!\left(\frac{t}{2}\right)+I_{-m}\!\left(\frac{t}{2}\right)
	\right],
\label{Bessel mod}
\end{equation}
\begin{equation}
	\int_{-\infty}^{\infty} du\; e^{-t\sinh^2 u}e^{2inu}
	=
	e^{t/2}K_{in}\!\left(\frac{t}{2}\right).
\end{equation}
Using only the first addend in \eqref{Bessel mod} and performing the same Mellin-transform steps as before, one obtains
\begin{equation}
	A(m,n)=
	C\,
	\frac{
	\Gamma\!\left(x+\frac{m+in}{2}\right)
	\Gamma\!\left(x+\frac{m-in}{2}\right)
	}{
	\Gamma\!\left(1-x+\frac{m+in}{2}\right)
	\Gamma\!\left(1-x+\frac{m-in}{2}\right)
	},
\end{equation}
where
\begin{equation}
	x=\frac{\Delta}{2},
	\qquad
	C=\pi\,2^{2\Delta-2}\frac{\Gamma(1-\Delta)}{\Gamma(\Delta)}.
\end{equation}
The full expression for $J(\Delta,m,n)$ is then obtained by adding the reflected contribution $m\to -m$, namely
\begin{equation}
	J(\Delta,m,n)=\frac{A(m,n)+A(-m,n)}{2}.
    \label{B6}
\end{equation}
This representation is manifestly real (since the Gamma functions appear in complex-conjugate pairs, both in the numerator and in the denominator).\\
It should be kept in mind that the integral representation of \(J(\Delta,m,n)\) is directly defined only for \(|m|<\Delta\). By analytic continuation, however, the Gamma-function expression extends it to the whole \(m\)-\(n\) plane, except at the poles located at \(n=0\) and \(m=\Delta+2k\), with \(k\in\mathbb N_+\), and analogously for negative \(m\), since the expression is even in both \(m\) and \(n\).

The same result can be written explicitly as
\begin{align}
J(\Delta,m,n)=\frac{\pi\,2^{2\Delta-3}\Gamma(1-\Delta)}{\Gamma(\Delta)}
\Bigg[&
\frac{\Gamma\!\left(\frac{\Delta+m+in}{2}\right)\Gamma\!\left(\frac{\Delta+m-in}{2}\right)}
{\Gamma\!\left(1-\frac{\Delta-m-in}{2}\right)\Gamma\!\left(1-\frac{\Delta-m+in}{2}\right)} \, +
\nonumber\\
&+
\frac{\Gamma\!\left(\frac{\Delta-m+in}{2}\right)\Gamma\!\left(\frac{\Delta-m-in}{2}\right)}
{\Gamma\!\left(1-\frac{\Delta+m-in}{2}\right)\Gamma\!\left(1-\frac{\Delta+m+in}{2}\right)}
\Bigg].
\label{eq:JGammaVersion1}
\end{align}
Equivalently, using the reflection formula for the modified Bessel function
\begin{equation}
I_{-\nu}(x)=I_\nu(x)+\frac{2}{\pi}\sin(\pi\nu)\,K_\nu(x)
\end{equation}
in \eqref{Bessel mod}, one obtains
\begin{align}
J(\Delta,m,n)= \
& \pi\,2^{2\Delta-2}\frac{\Gamma(1-\Delta)}{\Gamma(\Delta)}
\frac{\Gamma\!\left(\frac{\Delta+m+in}{2}\right)\Gamma\!\left(\frac{\Delta+m-in}{2}\right)}
{\Gamma\!\left(1-\frac{\Delta-m+in}{2}\right)\Gamma\!\left(1-\frac{\Delta-m-in}{2}\right)} \, + \nonumber \\
& + \sin(\pi m) \frac{2^{\,2\Delta-3}}{\Gamma^2(\Delta)} \Gamma\!\left(\tfrac{\Delta+m+i n}{2}\right)\,\Gamma\!\left(\tfrac{\Delta+m-i n}{2}\right)\,\Gamma\!\left(\tfrac{\Delta-m+i n}{2}\right)\,\Gamma\!\left(\tfrac{\Delta-m-i n}{2}\right)\, ,
\label{eq:JGammaVersion2}
\end{align}
where the second term vanishes for integer \(m\).

The parameter $m$ labels the spatial momentum sector, and therefore also the possible winding of the condensate around the spatial circle. In particular, for integer $m$ and setting $n=0$ in \eqref{eq:JGammaVersion2}, one finds
\begin{equation}
	J(\Delta,m,0)=
	\pi\,2^{2\Delta-2}\,
	\frac{\Gamma\!\left(\frac{\Delta+m}{2}\right)^2\Gamma(1-\Delta)}
	{\Gamma(\Delta)\,\Gamma\!\left(1-\frac{\Delta-m}{2}\right)^2}\,.
	\label{eq:Jm0_appendix}
\end{equation}
This leads to the new critical coupling
\begin{equation}
	f_d(m,0;L) = -\frac{1}{G_0(m,0;L)}=  \frac{1-\Delta}{\pi }
	\frac{\Gamma(\Delta-1)\,\Gamma\!\left(1-\frac{\Delta-m}{2}\right)^2}{\Gamma(1-\Delta)\Gamma\!\left(\frac{\Delta+m}{2}\right)^2}\,.
\end{equation}
This is the same expression as in the homogeneous case, up to the shift $m/2$ in the arguments of the squared Gamma functions. It can therefore be interpreted as the field theory counterpart of the winding soliton sector on the AdS side \cite{bolognesi}, where the same type of shift appears in the holographic coupling for non-zero winding modes \eqref{eq kappa soliton} (up to the different notation in which the AdS winding number is denoted by $n$ rather than $m$).

\section{Computation of the dynamical coefficients}
\label{app:coefficients}

In this appendix we derive the leading momentum dependence of the susceptibility by promoting the discrete mode labels to continuous variables and then taking the small-momentum limit
\begin{equation}
	m\to 0,
	\qquad
	n\to 0.
\end{equation}
This approximation is valid for momenta much smaller than the inverse cylinder scales. \\
Starting from \eqref{B6}, the logarithm of the Gamma function can be expanded as
\begin{equation}
	\ln\Gamma(z+\varepsilon)
	=
	\ln\Gamma(z)+\varepsilon\,\psi(z)
	+\frac{\varepsilon^2}{2}\psi_1(z)
	+\mathcal O(\varepsilon^3),
\end{equation}
where $\psi$ and $\psi_1$ are the digamma and trigamma functions, respectively. Let us then introduce
\begin{equation}
	f(\varepsilon)\equiv
	\ln\Gamma(x+\varepsilon)-\ln\Gamma(1-x+\varepsilon)
	=
	f(0)+\varepsilon D+\frac{\varepsilon^2}{2}E+\mathcal O(\varepsilon^3),
\end{equation}
with
\begin{equation}
	D=\psi(x)-\psi(1-x),
	\qquad
	E=\psi_1(x)-\psi_1(1-x).
\end{equation}
If we now define
\begin{equation}
	\varepsilon_1=\frac{m+in}{2},
	\qquad
	\varepsilon_2=\frac{m-in}{2},
\end{equation}
we have
\begin{equation}
	\varepsilon_1+\varepsilon_2=m,
	\qquad
	\varepsilon_1^2+\varepsilon_2^2=\frac{m^2-n^2}{2},
\end{equation}
and therefore
\begin{equation}
	A(m,n)=C\,\exp\!\bigl(f(\varepsilon_1)+f(\varepsilon_2)\bigr).
\end{equation}
Using also
\begin{equation}
	e^{2f(0)}=
\left[\frac{\Gamma(x)}{\Gamma(1-x)}\right]^2,
\end{equation}
one finds
\begin{equation}
	A(m,n)
	=
	J(\Delta)\left[
	1+mD+\frac{m^2-n^2}{4}E+\frac{m^2}{2}D^2
	+\mathcal O(\varepsilon^3)
	\right].
\end{equation}
Averaging with $m\to -m$, the linear terms cancel and we finally obtain
\begin{equation}
	J(\Delta,m,n)
	=
	J(\Delta)\left[
	1
	+m^2\left(\frac{E}{4}+\frac{D^2}{2}\right)
	-\frac{n^2}{4}E
	\right]
	+\mathcal O(m^4,m^2n^2,n^4).
    \label{J completo}
\end{equation}
This shows explicitly that the susceptibility is quadratic in the spatial and temporal momenta, as assumed in the derivative expansion of the effective action.

The result can be simplified further by using the standard identities
\begin{equation}
	\psi(1-z)-\psi(z)=\pi\cot(\pi z),
\end{equation}
which implies
\begin{equation}
	D=-\pi\cot(\pi x),
    \label{espressione D}
\end{equation}
and
\begin{equation}
	\psi_1(1-z)+\psi_1(z)=\pi^2\csc^2(\pi z),
\end{equation}
from which one gets
\begin{equation}
	E=2\psi_1(x)-\pi^2\csc^2(\pi x).
        \label{espressione E}
\end{equation}

To relate this result to the local derivative expansion of the effective action \eqref{eq:Seff_complete}, we match the small-momentum expansion of the susceptibility given in \eqref{G espanso} with the expansion obtained above.
Using \eqref{J completo} together with
\begin{equation}
    k_m=\frac{2\pi m}{L},
    \qquad
    \omega_n=\frac{2\pi n}{\beta},
\end{equation}
we find
\begin{equation}
    G_0(q)=G_0(0)\left[
    1+\left(\frac{E}{4}+\frac{D^2}{2}\right)\frac{L^2 k_m^2}{4\pi^2}
    -\frac{E}{4}\frac{\beta^2 \omega_n^2}{4\pi^2}
    +O(q^4)
    \right].
\end{equation}
Matching with \eqref{G espanso} gives
\begin{equation}
    c_x=\frac{G_0(0)L^2}{16\pi^2}\left(E+2D^2\right),
    \qquad
    c_\tau=-\frac{G_0(0)\beta^2}{16\pi^2}E.
\end{equation}
Using \eqref{espressione D} and \eqref{espressione E}, one may rewrite them as
\begin{equation}
    c_x=\frac{G_0(0)L^2}{16\pi^2}
    \left[
    2\psi_1(x)-\pi^2\csc^2(\pi x)+2\pi^2\cot^2(\pi x)
    \right], \quad c_\tau=\frac{G_0(0)\beta^2}{16\pi^2}
    \left[
    \pi^2\csc^2(\pi x)-2\psi_1(x)
    \right].
\end{equation}

The quadratic effective action is therefore not Lorentz invariant in general, since \(c_\tau\neq c_x\). Equivalently, the quadratic kernel is not a function only of the Lorentz-invariant combination \(-\omega^2+k^2\) (or of \(\omega_n^2+k_m^2\) in Euclidean signature). This is expected because the background is \(S^1_\beta\times S^1_L\), rather than the vacuum on \(\mathbb{R}^{1,1}\): the thermal circle selects a preferred frame and the compact spatial circle introduces the additional scale \(L\). Hence boosts mixing time and space are not symmetries of the background, and different coefficients for \((\partial_\tau\sigma)^2\) and \((\partial_x\sigma)^2\) are allowed. Lorentz invariance is recovered only in the decompactified zero-temperature limit, in which both compactification scales are removed.

\printbibliography[heading=bibintoc, title={References}]


\end{document}